\documentclass[twoside,twocolumn,9pt]{article}
\usepackage{gensymb}
\usepackage{extsizes}
\usepackage[super,sort&compress,comma]{natbib} 
\usepackage[version=3]{mhchem}
\usepackage[left=1.5cm, right=1.5cm, top=1.785cm, bottom=2.0cm]{geometry}
\usepackage{balance}
\usepackage{widetext}
\usepackage{times,mathptmx}
\usepackage{sectsty}
\usepackage{graphicx} 
\usepackage{caption}
\usepackage{subcaption}
\usepackage{lastpage}
\usepackage[format=plain,justification=raggedright,singlelinecheck=false,font={stretch=1.125,small,sf},labelfont=bf,labelsep=space]{caption}
\usepackage{float}
\usepackage{fancyhdr}
\usepackage{fnpos}
\usepackage[english]{babel}
\usepackage{array}
\usepackage{droidsans}
\usepackage{charter}
\usepackage[T1]{fontenc}
\usepackage[usenames,dvipsnames]{xcolor}
\usepackage{setspace}
\usepackage[compact]{titlesec}

\usepackage{epstopdf}

\definecolor{cream}{RGB}{222,217,201}

\begin{document}

\pagestyle{fancy}
\thispagestyle{plain}
\fancypagestyle{plain}{

\fancyhead[C]{\includegraphics[width=18.5cm]{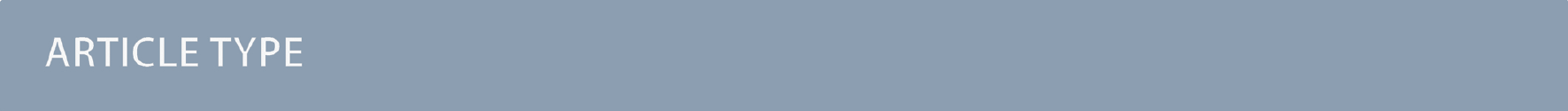}}
\fancyhead[L]{\hspace{0cm}\vspace{1.5cm}\includegraphics[height=30pt]{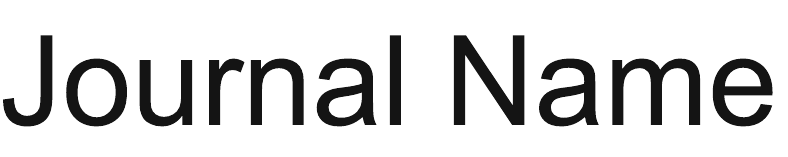}}
\fancyhead[R]{\hspace{0cm}\vspace{1.7cm}\includegraphics[height=55pt]{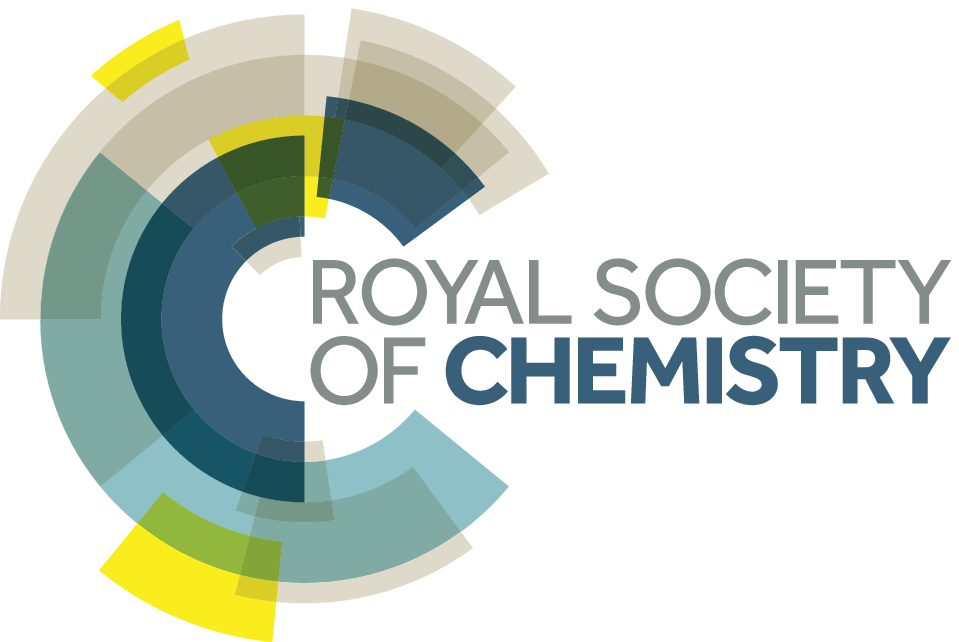}}
\renewcommand{\headrulewidth}{0pt}
}

\makeFNbottom
\makeatletter
\renewcommand\LARGE{\@setfontsize\LARGE{15pt}{17}}
\renewcommand\Large{\@setfontsize\Large{12pt}{14}}
\renewcommand\large{\@setfontsize\large{10pt}{12}}
\renewcommand\footnotesize{\@setfontsize\footnotesize{7pt}{10}}
\makeatother

\renewcommand{\thefootnote}{\fnsymbol{footnote}}
\renewcommand\footnoterule{\vspace*{1pt}%
\color{cream}\hrule width 3.5in height 0.4pt \color{black}\vspace*{5pt}} 
\setcounter{secnumdepth}{5}

\makeatletter 
\renewcommand\@biblabel[1]{#1}            
\renewcommand\@makefntext[1]%
{\noindent\makebox[0pt][r]{\@thefnmark\,}#1}
\makeatother 
\renewcommand{\figurename}{\small{Fig.}~}
\sectionfont{\sffamily\Large}
\subsectionfont{\normalsize}
\subsubsectionfont{\bf}
\setstretch{1.125} 
\setlength{\skip\footins}{0.8cm}
\setlength{\footnotesep}{0.25cm}
\setlength{\jot}{10pt}
\titlespacing*{\section}{0pt}{4pt}{4pt}
\titlespacing*{\subsection}{0pt}{15pt}{1pt}

\fancyfoot{}
\fancyfoot[LO,RE]{\vspace{-7.1pt}\includegraphics[height=9pt]{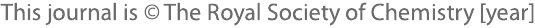}}
\fancyfoot[CO]{\vspace{-7.1pt}\hspace{13.2cm}\includegraphics{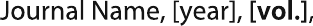}}
\fancyfoot[CE]{\vspace{-7.2pt}\hspace{-14.2cm}\includegraphics{RF}}
\fancyfoot[RO]{\footnotesize{\sffamily{1--\pageref{LastPage} ~\textbar  \hspace{2pt}\thepage}}}
\fancyfoot[LE]{\footnotesize{\sffamily{\thepage~\textbar\hspace{3.45cm} 1--\pageref{LastPage}}}}
\fancyhead{}
\renewcommand{\headrulewidth}{0pt} 
\renewcommand{\footrulewidth}{0pt}
\setlength{\arrayrulewidth}{1pt}
\setlength{\columnsep}{6.5mm}
\setlength\bibsep{1pt}

\makeatletter 
\newlength{\figrulesep} 
\setlength{\figrulesep}{0.5\textfloatsep} 

\newcommand{\topfigrule}{\vspace*{-1pt}%
\noindent{\color{cream}\rule[-\figrulesep]{\columnwidth}{1.5pt}} }

\newcommand{\botfigrule}{\vspace*{-2pt}%
\noindent{\color{cream}\rule[\figrulesep]{\columnwidth}{1.5pt}} }

\newcommand{\dblfigrule}{\vspace*{-1pt}%
\noindent{\color{cream}\rule[-\figrulesep]{\textwidth}{1.5pt}} }

\makeatother

\twocolumn[
\begin{@twocolumnfalse}
\vspace{3cm}
\sffamily
\begin{tabular}{m{4.5cm} p{13.5cm} }

\includegraphics{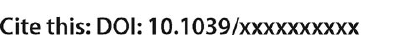} & \noindent\LARGE{\textbf{Quantitative Morphological Optimization of Bicontinuous Pickering Emulsions via Interfacial Curvatures$^\dag$}} \\
\vspace{0.3cm} & \vspace{0.3cm} \\

 & \noindent\large{Matthew Reeves$^{a}$ and Job H. J. Thijssen\textit{$^{a\ast}$}} \\

\includegraphics{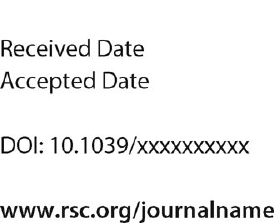} & \noindent\normalsize{Bicontinuous Pickering emulsions (bijels) are a physically interesting class of soft materials with many potential applications including catalysis, microfluidics and tissue engineering.  They are created by arresting the spinodal decomposition of a partially-miscible liquid with a (jammed) layer of interfacial colloids.  Porosity $L$ (average interfacial separation) of the bijel is controlled by varying the radius ($r$) and volume fraction ($\phi$) of the colloids ($L \propto r/\phi$).  However, to optimize the bijel structure with respect to other parameters, e.g. quench rate, characterizing by $L$ alone is insufficient.  Hence, we have used confocal microscopy and X-ray CT to characterize a range of bijels in terms of local and area-averaged interfacial curvatures.  In addition, the curvatures of bijels have been monitored as a function of time, which has revealed an intriguing evolution up to 60 minutes after bijel formation, contrary to previous understanding.} \\

\end{tabular}

 \end{@twocolumnfalse} \vspace{0.6cm}

  ]

\renewcommand*\rmdefault{bch}\normalfont\upshape
\rmfamily
\section*{}
\vspace{-1cm}


\footnotetext{\textit{$^{a}$SUPA School of Physics and Astronomy, James Clerk Maxwell Building, Peter Guthrie Tait Road, King's Buildings, Edinburgh, EH9 3FD, UK. Tel: +44 (0)131 650 5274; E-mail:  j.h.j.thijssen@ed.ac.uk}}

\footnotetext{\dag~The data corresponding to this paper will be made available at [DOI from Edinburgh Datashare]}




\section{Introduction}
Bijels (bicontinuous interfacially-jammed emulsion gels) are a versatile and relatively new class of soft materials, whereby a bicontinuous interlocking structure of two phase-separated liquids is stabilized by a (jammed) layer of interfacial colloids.\cite{Stratford2005,Cates2008,Firoozmand2015,Cui2013,Bai2015,Haase2015,Tavacoli2015} As long as the colloids are sufficiently neutrally-wetting, i.e. have no preference for either liquid phase, the bicontinuous pattern formed by the liquid-liquid (L-L) interface during spinodal decomposition can be maintained and eventually locked in (Figure 1). This is because, during the phase separation, the area fraction of particles increases until they jam, at which point the system stops coarsening.\cite{Herzig2007}  The bijel lends itself to a host of technological applications, including tissue engineering,\cite{Martina2005} catalysis\cite{Sung2005} and microfluidics,\cite{Tanaka2002} owing to its morphology and high degree of structural tunability.\cite{Stratford2005,Cates2008}   

The water/lutidine (W/L) bijel has been shown to be a particularly promising system for material templating \textendash{} one of the fluid phases can be selectively polymerized and, on removal of the unpolymerized liquid phase, results in a bicontinuous microporous monolith.\cite{Lee2010,Lee2012}  To tune the final structure of the material, samples are usually (and simply) characterized by their porosity, or rather their average interfacial separation (\textquoteleft{}channel width\textquoteright{}) $L$ (see Figure 1).  $L$ is controlled by varying the volume fraction ($\phi$) and radius ($r$) of the colloids ($L \propto r/\phi$) in the initial bijel mixture.  $L$ can be measured by constructing a pixel-based radial distribution function (in real space) from a 2D fluorescent confocal microscopy image of a bijel,\cite{Reeves2015a} in which one phase is bright and the other dark (owing to the preference of a fluorescent dye for one of the phases). Alternatively, the dominant wavevector $q^{\ast}$ can be determined by performing a scattering experiment\cite{Hashimoto1986} or by taking the Fourier transform of a 2D bijel image\cite{Herzig2007} \textendash{} then $L=2\pi/q^{\ast}$.  $L$ in the polymerized case can be extracted from X-ray micro-CT data in a similar fashion, where the contrast is provided by the difference in electron density across the solid/air interface.  

Measuring $L$ is relatively straightforward, but does not convey information about local geometrical or global topological parameters which might be useful for optimizing the bijel for applications.  Indeed, visible changes in the bijel morphology are noticed when different particle sizes and/or quench rates are used in the fabrication which seem not to be quantified purely by $L$.\cite{Reeves2015a}  Moreover, bijel fabrications produce samples which show a sliding scale of quality (as a function of particle size and quench rate), with complete failure on one end and homogenous \textquoteleft{}clean\textquoteright{} samples on the other.\cite{Reeves2015a}  It would be beneficial to have a quantitative description to accompany these qualitative observations, and hence be able to systematically optimize bijel fabrication procedures.

A complimentary way to characterize complex 3D structures such as the bijel is to measure the distributions of interfacial curvatures.\cite{Jinnai2001,Lee2010}  At each point on the interface between the coexisting liquid domains, it is possible to define two mutually-orthogonal principle radii of curvature (see Figure 1), $R_1$ \& $R_2$, and two principle curvatures $\kappa_1$ \& $\kappa_2$ ($\kappa_i = \pm 1/R_i$).\cite{Hyde1997a}  In the current study, the curvature is defined to be positive if the interface curves towards the oil phase, and negative if it curves towards the water phase.  The structure can then be characterized\cite{Hyde1997} by computing the area-averaged mean curvature:
\begin{equation}
\langle H \rangle = \int_A \frac{1}{2} (\kappa_1 + \kappa_2) \textrm{d}A
\end{equation}
and Gaussian curvature:
\begin{equation}
\langle K \rangle = \int_A \kappa_1 \kappa_2 \textrm{d}A
\end{equation}
This characterization allows the bijel's topology to be probed, as opposed to simply extracting a characteristic length ($L$).  

\begin{figure}[h]
\centering
\includegraphics[width=0.4\textwidth]{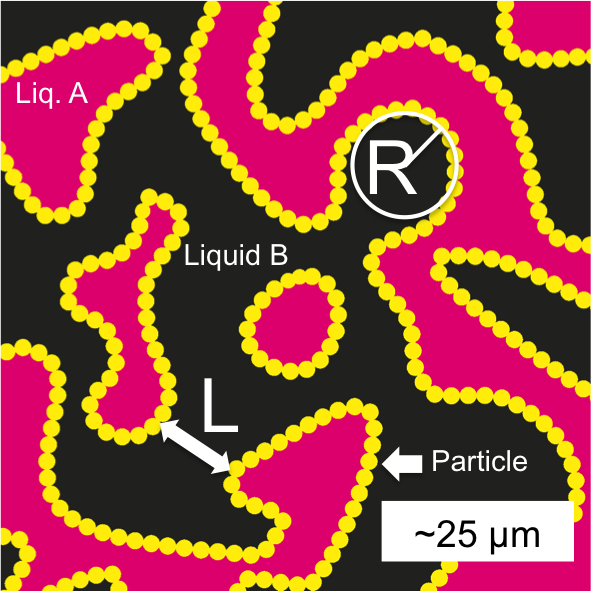}
\caption{A schematic of a 2D slice of a 3D bijel. The liquid phases (A and B) are separated by a jammed layer of particles (yellow) at the interface, with a typical separation $L$.  The principal radius of curvature $R$ at a point on the interface is the radius of a circle which intersects tangentially at that point, with a curvature $\kappa$ = 1/$R$.  In 3D a point will have two principal curvatures. Adapted from Reeves \textit{et al.}\cite{Reeves2015a}}
\label{curvatures_example}
\end{figure}

In addition to the practical motivations for the present study, there are several interesting physics questions to be explored.  For example, it is not completely clear whether (neutrally-wetting) particles simply \textquoteleft{}lock in\textquoteright{} the spinodal pattern of the demixing liquids, or if they perturb it in some way before arresting the phase separation.\cite{Cheng2013,Reeves2015a}  Any perturbation could also depend on other parameters like particle size and/or contact angle.\cite{Kralchevsky2005,Reeves2015a}  Moreover, in the W/L system, the interfacial particles are known to form a permanent gel after a certain incubation period\cite{Sanz2009} \textendash{} this may or may not have an effect on the morphology.  Finally, it is also not known yet if the bijel structure is entirely homogeneous and periodic. 

A (small-scale) curvature analysis on a W/L bijel and its polymerized counterpart has already been performed, on one size of particle and one quench rate only, which showed a distribution of mean curvature centred (approximately) on zero, and a negative value for the Gaussian curvature, which are the hallmarks of a member of the family of triply-periodic minimal surfaces, the gyroid.\cite{Lee2010}  This means that the L-L interface is, on average, composed of saddle points, with principal curvatures equal in magnitude but opposite in sign.  Bicontinuous systems similar in structure to bijels have also been analysed in this way.  For example, it has been shown that a spinodally-decomposing polymer blend also resembles the gyroid.\cite{Jinnai2001} More recently, a topological analysis has elucidated the mechanisms at play during the phase separation of the polymers.\cite{Saito2009}  

In this paper, we systematically characterize a range of W/L bijel samples by using 3D confocal microscopy data and a commercially available image analysis package; we benchmark our analysis by using X-ray CT data (with improved statistics over confocal) and simulated data of idealized structures. We measure distributions of interfacial curvatures, and then compute area-averaged quantities in order to characterize the topology of the structures as a function of particle size, quench rate, and time after bijel formation.  We find that the most hyperbolic samples are formed when using the smallest particles and the fastest quench rates, and that in some cases the distribution of local curvatures changes with time, which may be linked to the emergence of a particle-bonded network at the L-L interface.
\section{Experimental Methods}
\subsection{Particle synthesis}
The synthesis procedure for the particles used in this study has been previously reported.\cite{Reeves2015a}  The most relevant aspects are repeated here for the benefit of the reader.  St\"ober silica particles, labeled with fluorescein (or rhodamine) isothiocyanate (FITC or RITC) were synthesized with different radii by controlling the reaction temperature.  Sizes were obtained from dynamic light scattering (DLS) and transmission electron microscopy (TEM) measurements.  The three sizes used in this study are $r$ = 44 nm (TEM), 80/63 nm (DLS/TEM) and 367 nm (TEM) \textendash{} polydispersities (in terms of TEM results) were 5.5\%, 15.0\% and 5.5\% respectively.  The sizes obtained by TEM will be used in this study to provide a valid comparison.  The 367 nm and 63 nm particles were labelled with FITC, and the 44 nm particles were labelled with RITC.

Surface chemistry was designed to be similar by modifying the concentration of dye in the reaction mixture.  Neutral wetting conditions were achieved by drying small amounts ($\approx 50$ mg) at 170$\degree{}$C and 20 mBar pressure.  This process is known to remove surface bound water and may also cause moderate dehydroxylation of the silica surface.\cite{Zhuravlev2000}  Wetting properties were maintained across the range of particles used by tuning the drying time and inspecting confocal micrographs of the emulsion structures that they formed.  We denote the 44 and 63 nm particles as nanoparticles (NPs) and the 367 nm particles as microparticles (MPs).  

\subsection{Bijel preparation}
\begin{figure}[h]
\centering
\includegraphics[width=0.45\textwidth]{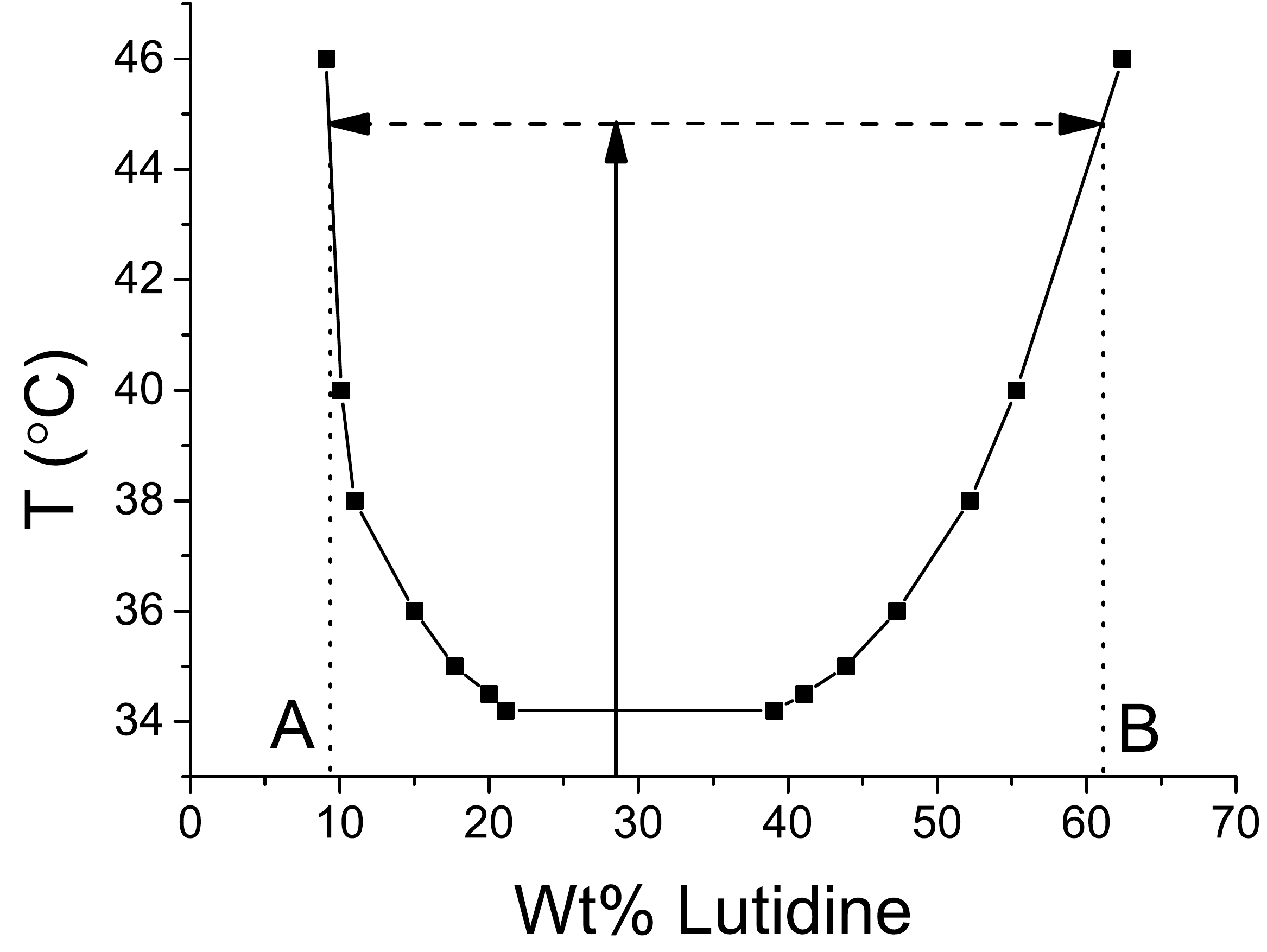}
\caption{Water/lutidine phase diagram, plotting existing data measured by Grattoni \textit{et al.}\cite{Grattoni1993}  Samples are prepared in the single-phase regime (i.e. below the critical temperature of 34.1\degree{}C) at a critical weight fraction of 28\% lutidine and are subsequently quenched to 45\degree{}C (or sometimes 50\degree{}C).  Coexisting phases of lutidine-poor (composition marked A) and lutidine-rich (composition marked B) are created.}
\label{Phase_diagram}
\end{figure}

Bijels were prepared by first dispersing the MPs or NPs in deionized water by ultrasonication (Sonics Vibracell) emulating a procedure used in a previous bijel study.\cite{Reeves2015a}  The MPs were sonicated for (2 $\times$ 2) minutes at 8W power, with 10 seconds of vortex mixing after each sonication step.  The NPs were sonicated for an extra 10 minutes at 8W with an extra 10 seconds of vortex mixing, to ensure proper redispersion.  Subsequently, lutidine (2,6-dimethylpyridine, $\geq$99\% Aldrich), was added along with Nile Red dye (or Nile Blue if the RITC labelled NPs were to be used) at a concentration of around 10 $\mu$M, to give a mixture with mass ratio of W:L = 72:28. This particular mass ratio was used to prepare the system at its critical composition,\cite{Grattoni1993} so that spinodal decomposition would be (at least initially) the preferred phase separation mechanism.\cite{Herzig2007}  Mixtures were transferred to a glass cuvette (1 mm path length, Starna Scientific) and placed inside a metal block, which was itself placed inside a temperature stage (Instec). The mixtures were quenched from room temperature up to 45 or 50\degree{}C (the vertical line in Figure \ref{Phase_diagram}) in various ways depending on the desired rate.  A rate of 1\degree{}C/min was achieved by programming the temperature stage \textendash{} a thermocouple was used to verify that the sample temperature did not lag the stage temperature.  A rate of 17\degree{}C/min was achieved by pre-warming the metal block to 45\degree{}C before the insertion of the cuvette.  A rate of 350\degree{}C/min was achieved by placing the cuvette on the top of a small cardboard box inside a microwave (DeLonghi, P80D20EL-T5A/H, 800W), set to \textquoteleft{}auto-defrost\textquoteright{} for 5 or 6 seconds.  In all cases, when the temperature exceeded 34.1\degree{}C the mixture phase separated via spinodal decomposition, resulting in the particles attaching to the liquid-liquid interface and ultimately arresting the bicontinuous structure, with coexisting phase compositions given by the positions A \& B in Figure \ref{Phase_diagram}.  Some bijels were polymerized following procedures described by Lee \& Mohraz.\cite{Lee2010}
\subsection{Data acquisition}
\begin{figure}[h]
\centering
\includegraphics[width=0.45\textwidth]{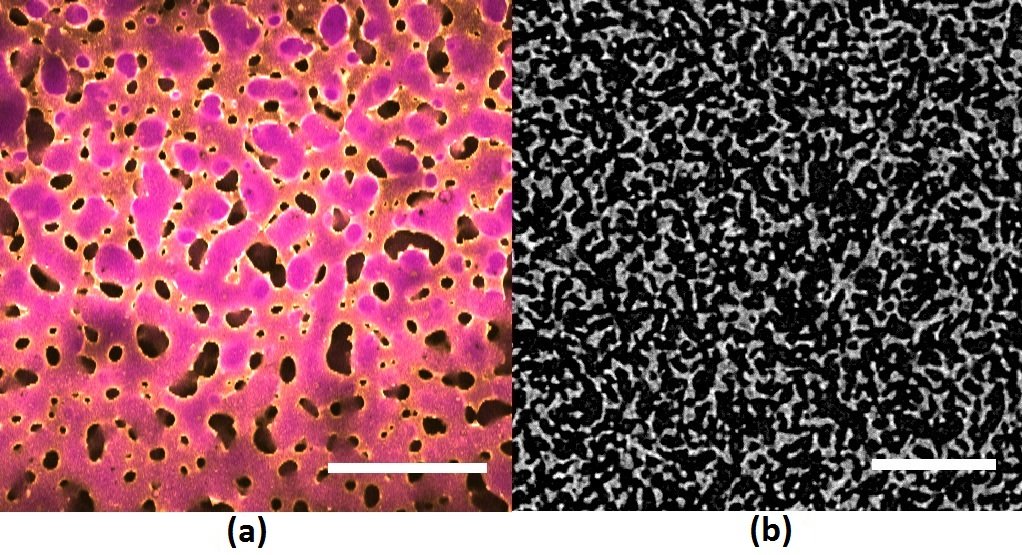}
\caption{(a) Example fluorescence confocal microscopy image of a bijel stabilized by NPs ($r=$ 63 nm), quenched at 17$\degree$C/min.  Particles are shown in yellow, whereas magenta indicates the lutidine-rich phase.  Scale bar 200$\mu$m.  Z-stacks are compiled by acquiring these 2D (X-Y) images at incrementing Z positions.  (b) Example X-ray CT image of a polymerized bijel, where one of the channels is air (black) and the other polymer (white).  Scale bar 500$\mu$m.  Imaging with this method allows a larger sample volume to be probed, providing better statistics than confocal microscopy.}
\label{Ex_conf}
\end{figure}

The bijel samples were imaged using fluorescence confocal microscopy, specifically a Zeiss Observer.Z1 inverted microscope with a Zeiss LSM700 confocal scanning unit.  Images of dimension 512 $\times$ 512 pixels were taken with a 20$\times$ or 40$\times$ objective (Zeiss LD-Plan NEOFLUAR, NA = 0.4 and 0.6 respectively), giving a pixel size of 1.25 $\mu$m or 0.625 $\mu$m.  Figure \ref{Ex_conf}(a) shows an example 2D confocal microscopy image of a bijel.  3D stacks were created by taking up to 60 images in succession with a (nominal) increment in the z-direction of about 1 $\mu$m per slice.  The acquisition of a 3D stack took between 6 and 8 minutes.  Fluorescence excitation was provided by a 488 nm laser (for FITC), a 555 nm laser (for Nile Red or RITC),  and a 639 nm laser (for Nile Blue).  Emission filters were used as appropriate. The two liquid domains could be distinguished by detecting the fluorescence of the Nile Red (or Blue) which preferentially resides in the lutidine-rich phase.  To take account of the decrease in image intensity as the focal plane is moved further in to the sample, the histograms of the 3D stacks were equalized (using the stack histogram) in ImageJ version 1.49.\cite{Rasband}

\subsection{Image analysis}
The 3D data was analyzed using Avizo (FEI Group).  An appropriate pixel value threshold was chosen to allow the software to generate an isosurface  via the \textquoteleft{}Isosurface\textquoteright{} module, which represents the interface between the two liquid domains.   The isosurface was \textquoteleft{}downsampled\textquoteright{} (averaged) over a volume of 3 $\times$ 3 $\times$ 3 pixels to reduce the impact of noise.  This produced a more accurate isosurface in terms of its resemblance to the original bijel (see Supplementary Information for an example), as well as reducing the occurrence of surface discontinuities (regions of non-manifold topological structure) on which the measurement of curvature becomes invalid. 

The software was then instructed to extract the distribution of mean ($H = \frac{1}{2}(\kappa_1 + \kappa_2)$) and Gaussian ($K = \kappa_1 \kappa_2$) curvatures \textendash{} $\kappa_1$ and $\kappa_2$ are the principal curvatures ($\kappa_i = \pm 1/R_i$, where $R_i$ are the radii of curvature) at each vertex on the triangulated isosurface.  The software achieves this by  approximating the isosurface locally (a region spanning 4 orders of connecting vertices in all directions, covering an area of a few $\mu \mathrm{m^2}$) in quadric form, then computing the eigenvalues and eigenvectors (the eigenvalues being the principal radii of curvature).  This function is contained within the \textquoteleft{}Compute curvatures\textquoteright{} module.  The computed curvatures were also averaged over a region spanning 5 orders of adjacent vertices, which reduced the impact of noise further.  These specific settings were chosen because they produced the results closest to that expected for our benchmark structure: a simulated gyroid with a similar lengthscale as our bijels in a data volume similar to our confocal stacks - see section 2.6.

Since $H$ and $K$ will naturally vary with the characteristic lengthscale of the structure\cite{Hyde1997} (smaller lengthscales neccesitate larger curvatures) it is necessary to normalize these quantities.  This can be done simply by multiplying by the inverse or inverse-square of the surface-to-volume ratio ($\Sigma = A/V$), which can also be extracted using the software.  The reduced dimensionless quantities are therefore $H \Sigma^{-1}$ and $K \Sigma^{-2}$.  

To reduce the potential impact of artefacts during the software's reconstruction of the bijel interface (i.e. regions of abnormally large curvature) only curvature values in a certain range were used to calculate the area-averaged quantities.  This range was chosen to be around the expected curvatures as derived from the measurement of $L$, i.e. $|H_\mathrm{expected}| \approx 2/L$ and $|K_\mathrm{expected}| \approx 4/L^2$.  Once the curvature distributions were truncated in this way (resulting in a loss of no more than 5\% of the total data), the values were summed to give $\langle H \rangle \Sigma^{-1}$ and $\langle K \rangle \Sigma^{-2}$.  Note that this is not the strict calculation of the integral of $H$ or $K$ over the surface required to apply the Gauss-Bonnet theorem,\cite{Hyde1997a} but rather the sum of a discrete series of values, without each value being weighted by the area over which the curvature was measured.  However, these values still serve as a (scaled) metric to quantify the bijel structure, since it is still sensitive to changes in the distributions.  For a bijel, we expect a structure with overall hyperbolic character, $\langle H \rangle \Sigma^{-1} \approx 0$, and $\langle K \rangle \Sigma^{-2}< 0$, although $\langle H \rangle \Sigma^{-1}$ may take a finite value close to zero if the volume ratio between the two phases is not 50:50,\cite{Hyde1997} as is the case in the W/L system employed here\cite{Grattoni1993} (see Figure \ref{Phase_diagram}).

\subsection{Errors}
Sampling errors have been accounted for by averaging the measurements of three uniquely prepared samples with identical procedures, although in some cases the bijel lengthscales were different.  Note that this also means that our results are unlikely to be significantly affected by small wetting differences between batches of dried particles.  Where repeat sampling was not feasible (e.g.  not enough identical material being available) the threshold value was varied around that which gave the maximum area, and an average of those curvature values taken.  Hence, the error bars on these data points represent the thresholding error in the measurement of one sample.  It will be made clear in the text which data points have been multiply sampled and which have not.  Further discussion of errors, including the fishtank effect, is contained within the Supplementary Information.

\subsection{Benchmarking}
\subsubsection{Simulated spheres and gyroid}
\begin{figure}[h]
\centering
\includegraphics[width=0.3\textwidth]{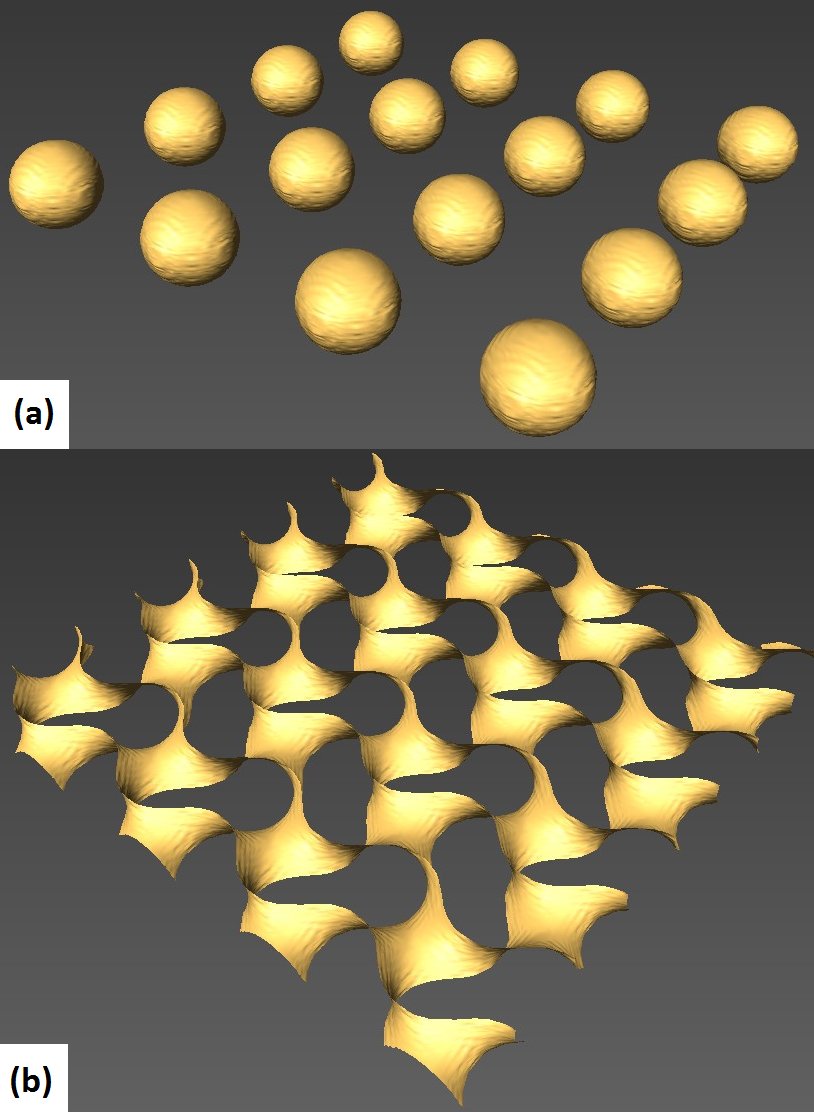}
\caption{Isosurfaces computed for benchmark structures, system of spheres (a) and a gyroid (b).  Matlab was used to numerically generate these structures (see text for details).}
\label{gyroid}
\end{figure}
In order to test the ability of the analysis method to accurately probe the topology of our bicontinuous structures, we generated idealized structures (representing the two possible extremes) with similar pixel sizes, lengthscales and box sizes as the real data.  On the one hand, a system of spheres was generated as shown in Figure \ref{gyroid}(a).  On the other, a gyroid structure (Figure \ref{gyroid}(b)) with roughly the same characteristic lengthscale and dimensions as the bijel data was generated by using the following equation;\cite{Nishikawa1998a,Matsushita1998}
\begin{equation}
f(x,y,z) = \sin \frac{2\pi x}{\lambda} \cos \frac {2\pi y}{\lambda} + \sin \frac{2\pi y}{\lambda} \cos \frac{2\pi z}{\lambda} + \sin \frac{2\pi z}{\lambda} \cos \frac{2\pi x}{\lambda}
\end{equation}
Here,  $\lambda$ is the characteristic wavelength, which in terms of bijel structure, is twice the interfacial separation $L$.  To create the dividing surface, $f(x,y,z)$ is set to zero.\cite{Jinnai1997}  To reiterate, we choose these structures because a bijel is expected to have $\langle H \rangle = 0$ and $\langle K \rangle < 0$ (i.e. the same as a gyroid), whereas a system of spheres has $\langle H \rangle >0$ and $\langle K \rangle >0$.

Table 1 shows the expected (E) and measured (M) values, along with the relative error (Err).  We concluded from the test results that we could use our analysis technique to accurately measure bijel topology.  We also explored the possible dependence of the accuracy on $\lambda$ at fixed box size but found that within the range of $L$ actually studied (in terms of bijels) this is not a concern (see Supplementary Information).  
\begin{table}
\begin{tabular}{c|c|c|}
Test & $\langle H \rangle \Sigma^{-1}$  E/M/Err & $\langle K \rangle \Sigma^{-2}$ E/M/Err \\
\hline
Spheres, $r=$ 32 pixels & 5.05/4.78/5.3\% & 25.5/22.9/10\% \\
Gyroid, $\lambda =$ 128 pixels & 0/0.00/0\% & -1.7/-1.66/2.4\%\\
\end{tabular}
\caption{The results of the analysis test on the benchmark structures shown in Figure \ref{gyroid}.  Expected (E) and measured (M) values are shown, along with a relative error (Err).}
\end{table}

\subsubsection{Data volumes}
The limitations of confocal microscopy in this case, namely the reduction in signal to noise (due to scattering) as the focal plane is moved further inside the sample, mean that acquired 3D data is constricted in the Z-direction.  The dimensions of the XY plane are 640 $\times$ 640 $\mu$m, whereas the Z dimension is only about 80 $\mu$m.  This may reduce the statistics when measuring interfacial curvatures, since there is clearly more information about 2 of the directions (XY) and less about the third (Z).  To see what effect this may have, we have acquired larger scale 3D data from polymerized bijels\cite{Lee2010} using X-ray computed tomography.    
The samples were micro-CT scanned using a Skyscan 1172 system (Bruker microCT, Kontich, Belgium). Rotating samples in angular steps of 0.4\degree{} over a full 360\degree{}, shadow-projection images with a pixel size of 4.48 $\mu$m were acquired (Hamamatsu 10 Mp camera).  The source voltage was 28 kV (Hamamatsu 100/250); no filter was used between source and sample, but a beam-hardening correction of 20\% was applied during reconstruction. Image reconstruction into axial slices was performed using the NRecon software package (Skyscan, version 1.5.1.5), yielding 3D data sets much larger than but similar in format to confocal data sets, i.e. Z-stacks of XY-slices. For reconstruction, a random movement of 10 was used, but no frame averaging was required to reduce the signal-to-noise ratio.

A comparison of the analyses of both types of data is presented in section 3.1.



\section{Results and discussion}
\subsection{Confocal vs Xray CT}
\begin{figure}[h]
\centering
\begin{subfigure}[b]{0.23\textwidth}
\includegraphics[width=\textwidth]{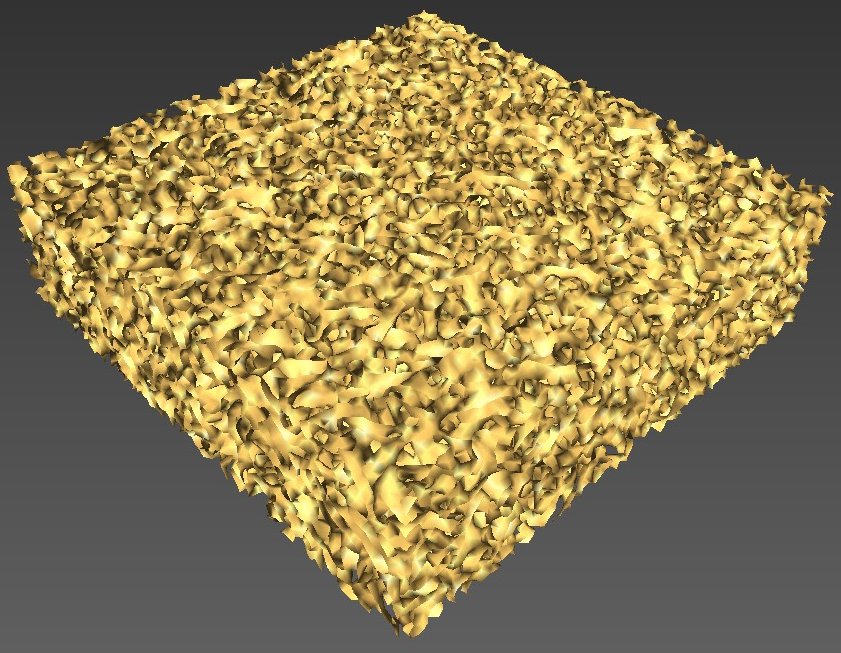}
\caption{}
\end{subfigure}
\begin{subfigure}[b]{0.23\textwidth}
\includegraphics[width=\textwidth]{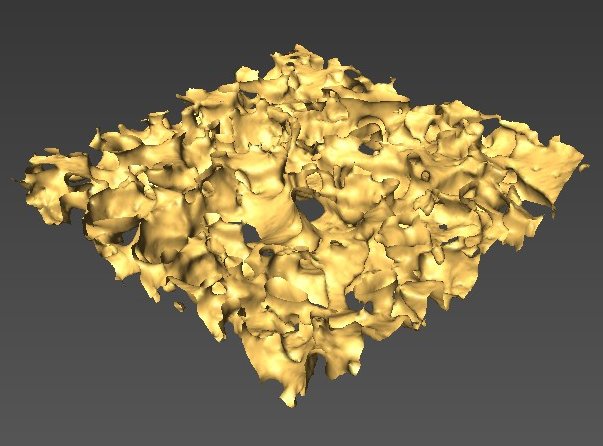}
\caption{}
\end{subfigure}
\begin{subfigure}[b]{0.23\textwidth}
\includegraphics[width=\textwidth]{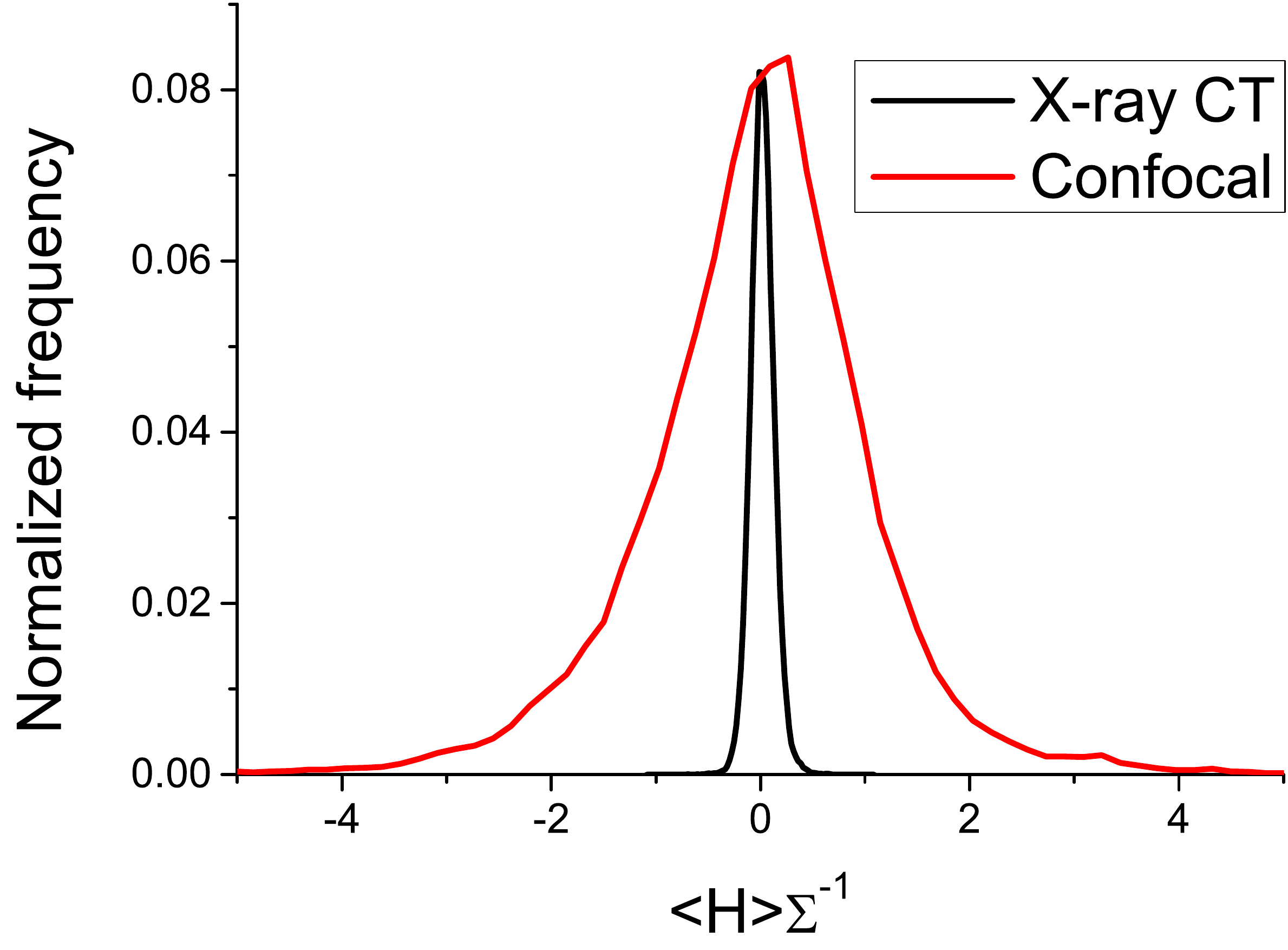}
\caption{}
\end{subfigure}
\begin{subfigure}[b]{0.23\textwidth}
\includegraphics[width=\textwidth]{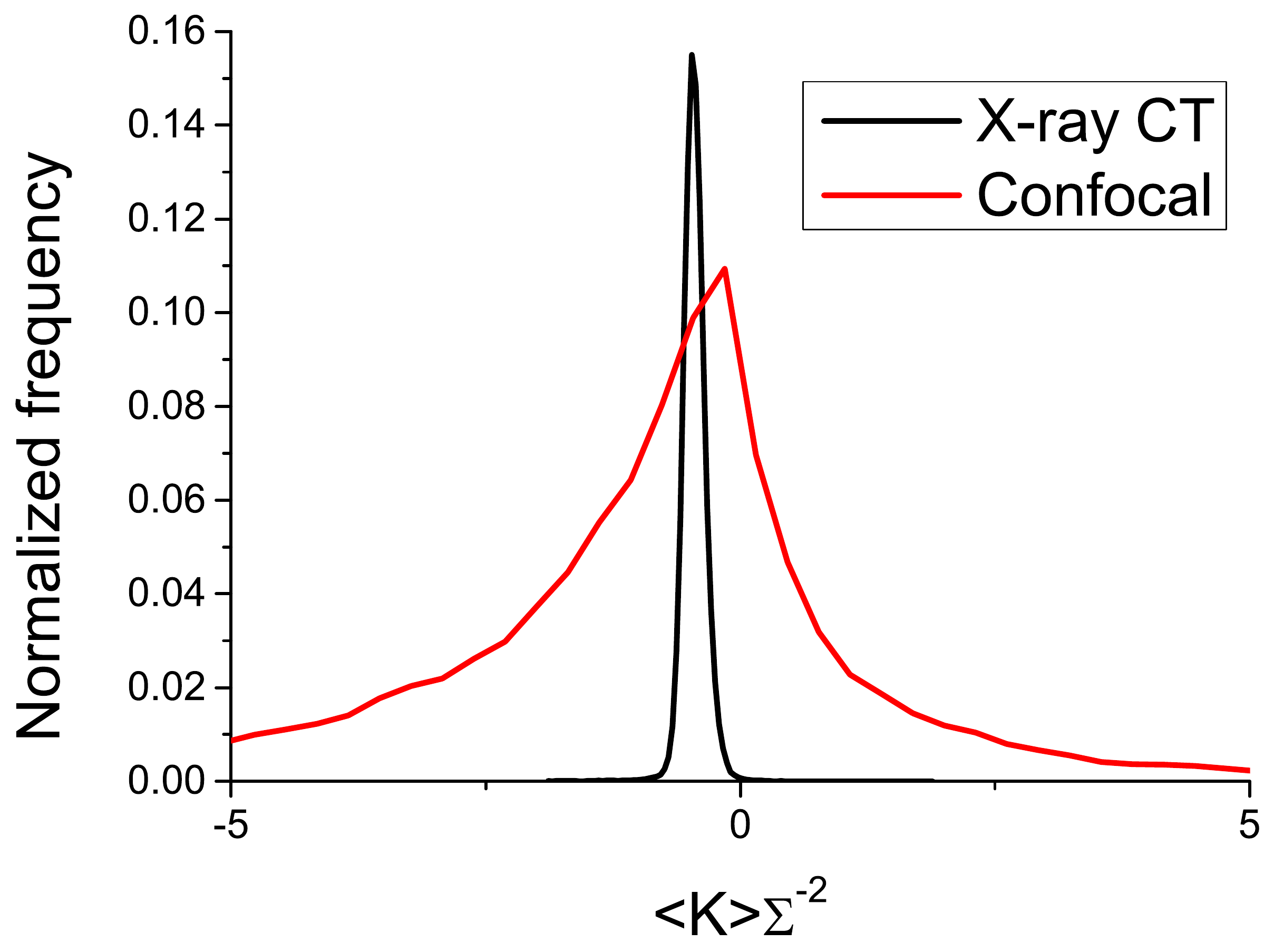}
\caption{}
\end{subfigure}
\caption{Example data used in the curvature analysis.  (a) Isosurface created using X-ray CT data. (b) Isosurface created using confocal data. (c) Mean curvature distributions for both isosurfaces. (d) Gaussian curvature distributions for both isosurfaces. The distributions were normalised by dividing the raw counts by total number of counts.}
\label{Comparison}
\end{figure}
Here we compare the analysis of data obtained from confocal microscopy (of a 63 nm NP stabilized bijel) and that obtained from X-ray CT (of a polymerized 63 nm NP stabilized bijel).  Figure \ref{Comparison} shows example isosurfaces and distributions of mean and Gaussian curvatures.  Since the X-ray method allows the sampling of much larger volumes than confocal, the isosurfaces created are much larger in area (Figure \ref{Comparison}(a) vs \ref{Comparison}(b)).  This allows the measurement of curvature over a larger number of bijel channels, increasing the accuracy.  The increased precision, evidenced by the sharper distributions in (c) and (d), is due to the sharper contrast between the channels, as the confocal data suffers the consequences of the point-spread function and multiple scattering, whereas the X-ray data does  not.  Although the distributions are sharper, the shape and position of the distributions are qualitatively similar, and produce similar area-averaged quantities: $\langle H \rangle \Sigma^{-1} = -0.02(1)$ and $\langle K \rangle \Sigma^{-2} = -0.8(1)$ in the case of the confocal data, and $\langle H \rangle \Sigma^{-1} = 0.002(5)$ and $\langle K \rangle \Sigma^{-2} = -0.45(1)$ in the case of the X-ray data.  

A comparison of the curvature distributions of a bijel and its polymerized counterpart has been made previously,\cite{Lee2010} showing $\langle H \rangle \approx 0$ in both cases (but with a small broadening in the negative region for the polymer) and $\langle K \rangle < 0$ in both cases (but with a small shift towards less negative values for the polymer).  The results presented in this study are in broad agreement with the results of Lee \textit{et al.}, with both mean curvatures approximately zero and a less negative value for the Gaussian curvature of the polymerized sample.  The difference in Gaussian curvature for the polymerized sample cannot be unequivocally explained by the polymerization process however, as the samples were prepared independently.   

Nonetheless, the fact that we get acceptably similar results when we use the method with a lower signal-to-noise ratio (confocal) allows us to use this in-situ characterization technique to study the effect of particle size and quench rate on bijel topology.

\subsection{Effect of particle size}
 
\begin{figure}[h]
\centering
\begin{subfigure}[b]{0.35\textwidth}
\includegraphics[width=\textwidth]{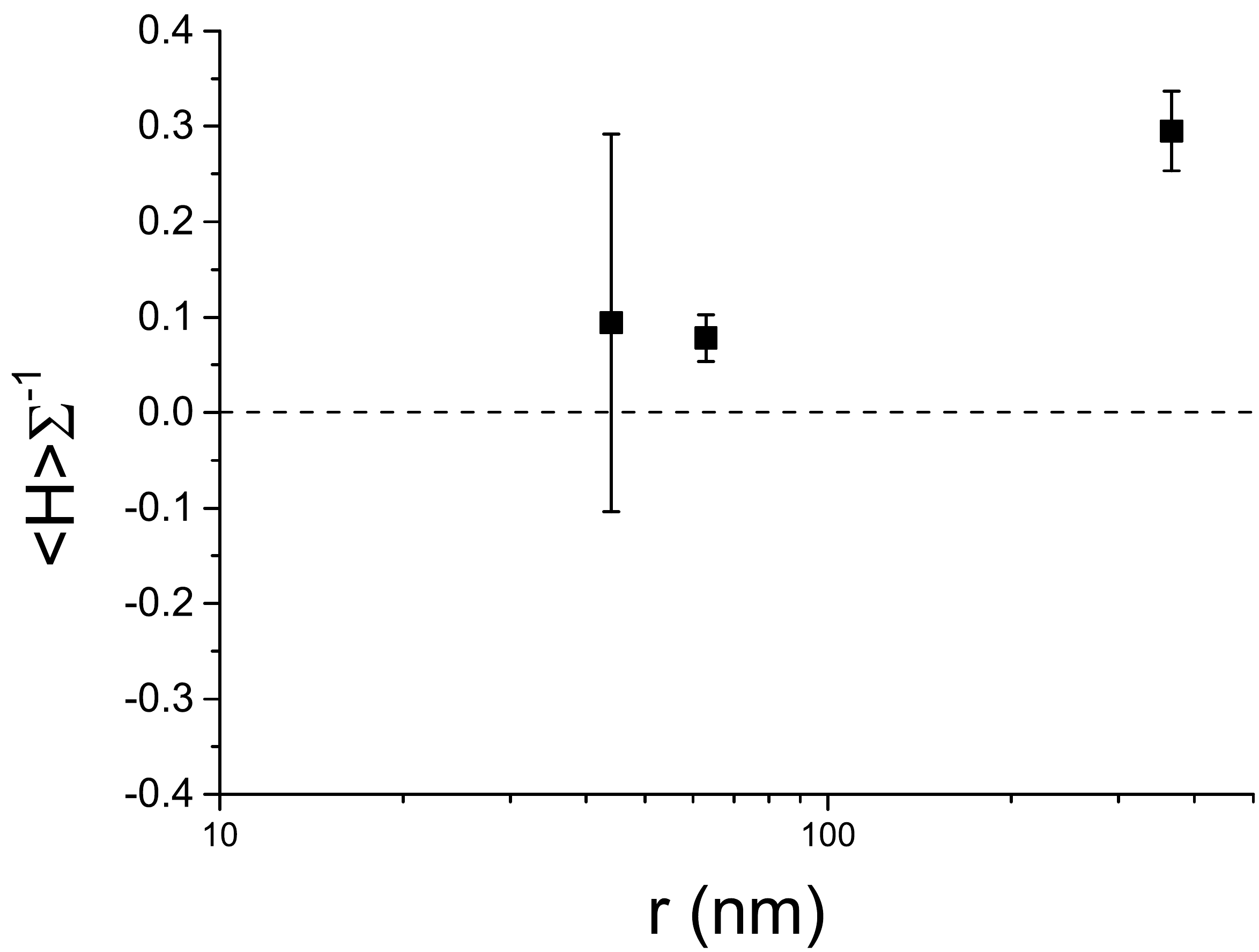}
\caption{}
\end{subfigure}
\begin{subfigure}[b]{0.35\textwidth}
\includegraphics[width=\textwidth]{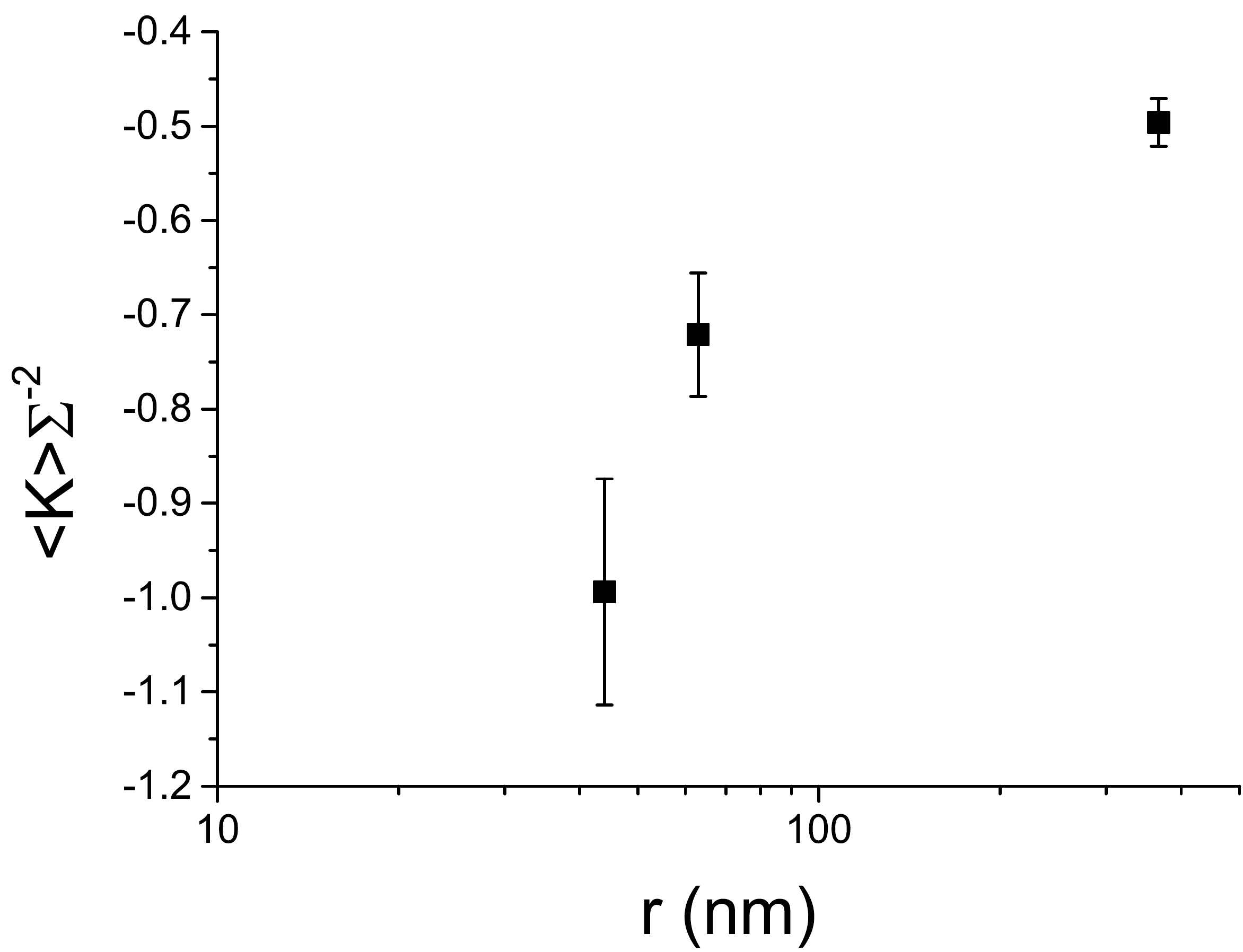}
\caption{}
\end{subfigure}
\caption{The change in area-averaged mean curvature (a) and Gaussian curvature (b) as a function of particle size.  The error bars for the two right-most points indicate one standard deviation of the distribution of 3 separate measurements, whereas the left-most point indicates the thresholding error from one unique measurement (for (a) and (b)).}
\label{particle_size}
\end{figure}
We now focus on data obtained from confocal microscopy, and vary the size of the particles used to stabilize the bijel, while keeping the quench rate constant at 350\degree{}C/min (microwave method).  The bijel lengthscales were in the range 31 $\mu \mathrm{m} < L < $ 91 $\mu \mathrm{m}$.  

Figure \ref{particle_size} shows how the area-averaged curvatures vary with particle size.  The data corresponding to $r = $ 63 and 367 nm has been averaged over three uniquely prepared samples, whereas the data for $r=$ 44 nm has only been sampled once \textendash{} hence, the error bars for the former indicate one standard deviation, and the error bar for the latter indicate the thresholding uncertainty.  

The area-averaged mean curvatures of the (final) bijel structures show a small dependence on particle size, with a significant positive value for the MP bijel and smaller positive values for the NPs.  This does not contradict the results of Lee \textit{et al.} since an inspection of the histograms reveals that all distributions are {\it centred} on zero.  These results are indicative of a (stronger) preferred direction of curvature for the MP bijel and a (weaker) preferred direction for the NP bijels.  

The area-averaged Gaussian curvatures show a clearer dependence on particle size \textendash{} the smaller the particle, the more negative the value.  In other words, as the size of the stabilizing particles is decreased, the bijel becomes more hyperbolic in character, and more closely resembles the ideal gyroid.  
 
These results can be partly explained by considering the following.  The batch of particles used in the bijel mixture will have a distribution of contact angles centred close to the desired 90\degree{},\cite{Isa2011} and the particles with $\theta \neq 90$\degree{} will induce a spontaneous curvature on the L-L interface.\cite{Kralchevsky2005}  Although NPs will induce a larger spontaneous curvature than MPs, the driving force towards this curvature is smaller.\cite{Reeves2015a}  Hence, it is expected that NPs are less disruptive to the L-L interface than MPs, and hence more likely to lock-in the spinodal structure of the L-L phase separation. The data presented here is largely in agreement with this.  The relative lack of preferred curvature direction for the NP stabilized bijels is consistent with the picture of the NPs being more successful at preventing pinch-off events, resulting in a larger number of (hyperbolic) fluid necks rather than (elliptic) bulging regions which would exist immediately after pinch-off.  


To assess how closely the bijels resemble the L-L interface without particles, we refer to previous work on a mixture of polybutadiene and poly(styrene-ran-butadiene) undergoing spinodal decomposition.\cite{Jinnai1997}  By using their measured data for the quantities $\Sigma / q_m = 0.5$ (where $q_m$ is the position of the peak in the Fourier intensity spectrum, i.e. after the 3D image data has been transformed into frequency space) and $\langle R \rangle / L = 0.3$ (where $\langle R \rangle$ is the average radius of curvature), we have calculated a value of $\langle K \rangle \Sigma^{-2}\approx -1.13$.  Note that this is significantly different from the ideal gyroid, but since the gyroid is a {\it minimal} surface ($H=0$ everywhere) it may be better at quantitatively describing {\it equilibrium} structures rather than the non-equilibrium structures reported here, for which we and others see a distribution of $H$ {\em centred} on 0.\cite{Nishikawa1998a,Lee2010}  Also, there are differences between the system studied here and the system studied by Jinnai \textit{et al.}, namely the ratio of interfacial tension to viscosity, so the calculated value is only an estimation.  However, it can still be concluded that the NPs lock-in structures with greater hyperbolic character which more closely resemble the L-L spinodal interface. 

From the point of view of applications, we now have a \textit{quantitative} characterization of bijel topology that allows rational optimization of synthesis parameters. For example, if one wanted to design a symmetric cross-flow microreactor from a bijel, then a hyperbolic rather than a parabolic surface would be desirable (to reduce the occurence of bottlenecks and/or dead-ends), and our analysis shows that using smaller particles would be beneficial. To further demonstrate this approach to bijel-materials design, we continue by considering two additional bijel-synthesis parameters: quench rate and age (i.e. time after bijel formation).

\subsection{Effect of quench rate}
\begin{figure}[h]
\centering
\begin{subfigure}[b]{0.4\textwidth}
\includegraphics[width=\textwidth]{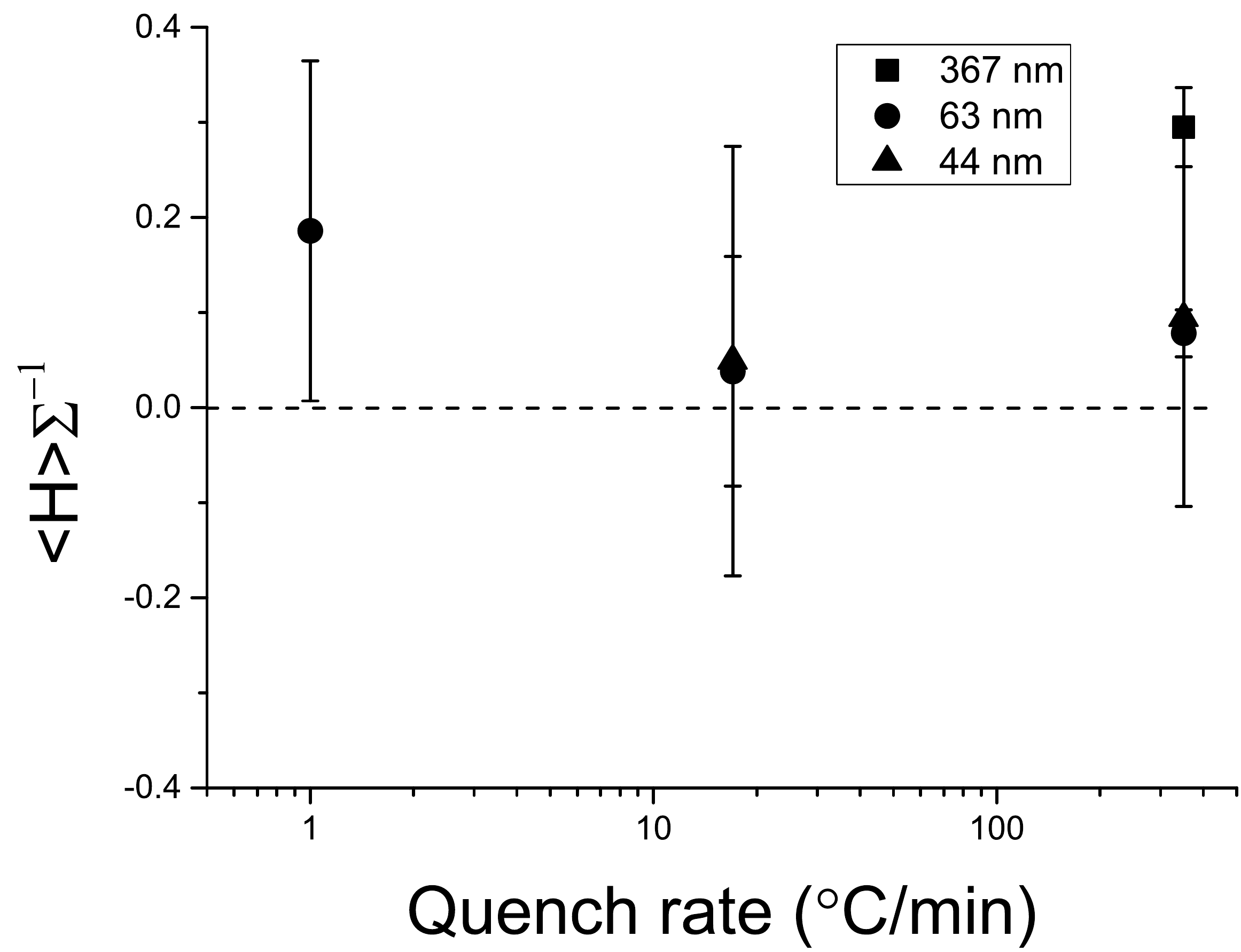}
\caption{}
\end{subfigure}
\begin{subfigure}[b]{0.4\textwidth}
\includegraphics[width=\textwidth]{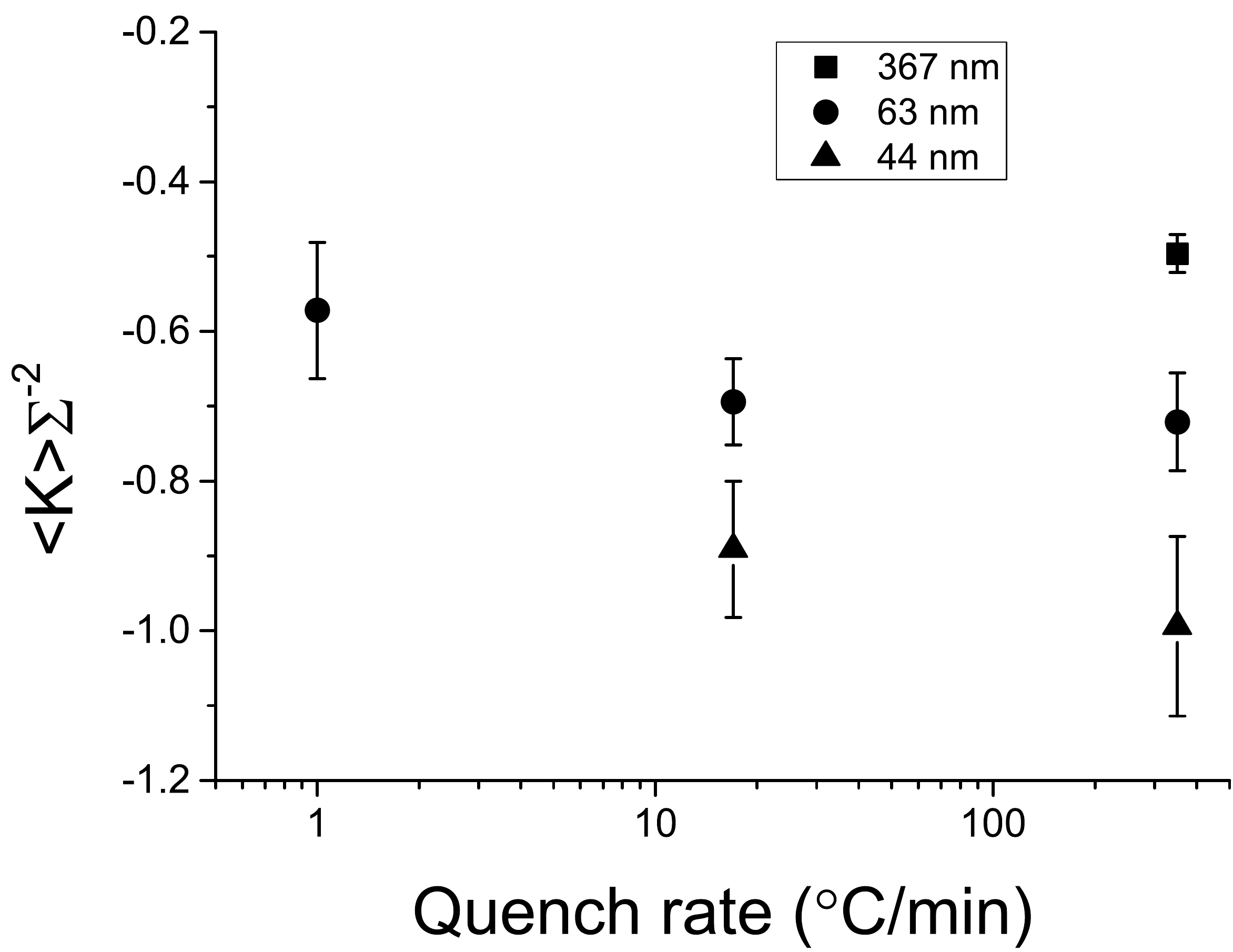}
\caption{}
\end{subfigure}
\caption{The change in area-averaged mean curvature (a), Gaussian curvature (b) as a function of quench rate.  The dashed line in (a) indicates the value expected for the ideal gyroid.}
\label{quench_rate}
\end{figure}
Here we continue to focus on confocal microscopy data, but now vary the quench rate used in the synthesis step (see Experimental Methods).  Note that in the bijel literature, the term \textquoteleft{}quench\textquoteright{} can refer to either cooling (in the case of an upper critical solution temperature, UCST) or heating (in the current case of a lower critical solution temperature, LCST).  Figure \ref{quench_rate} shows how the area-averaged curvatures vary with quench rate, with the 350\degree{}C/min data points being the same as shown in Figure \ref{particle_size} but on new axes.  The 44 nm data and the 1\degree{}C/min data has only been sampled once here.  The bijel lengthscales were in the range 21 $\mu$m $< L <$ 91 $\mu$m.

The area-averaged mean curvatures in the case of NP bijels remain constant upon the decrease to 17\degree{}C/min, but rise upon a further decrease to 1\degree{}C/min.  The preferred direction indicated by this data point could either be due to the presence of droplets (of a size similar to the channel width) in slowly-quenched samples due to secondary nucleation\cite{Witt2013} or due to the same mechanism which results in the preferred direction in (fast quenched) MP bijels. 

The area-averaged Gaussian curvatures show a slight variation with quench rate, becoming less hyperbolic as the quench rate is reduced.  Interestingly, it appears that quenching the 63 nm particle bijel at 1\degree{}C/min yields a structure similar to a 367 nm particle bijel, quenched at 350\degree{}C/min.  

To explore why this might be the case, we explain the effect of quench rate on the phase separation dynamics.  In the current case of spinodal decomposition, we assume that the relevant phase-separation regime is the \textquoteleft{}viscous hydrodynamic\textquoteright{} one (based on estimates of the timescale of the crossover between the diffusive and viscous hydrodynamic regimes).\cite{Pagonabarraga2001}  In this regime, the characteristic lengthscale of the structure scales with time according to $L \propto \frac{\gamma}{\eta}t$, where $\gamma$ is the interfacial tension and $\eta$ the fluid viscosity;  $\gamma$ scales with the quench depth as $\Delta T^{0.88}$, which itself scales with time according to the quench rate used ($\Delta T = \beta t$, where $\beta$ is the quench rate, and t = 0 when the temperature reaches 34.1\degree{}C).  $\eta$ shows some variation in the relevant temperature range but for the purposes of this illustration can be approximated as constant.  Hence, varying the quench rate changes the prefactor in the growth law, which is akin to saying that it changes the phase separation \textquoteleft{}speed\textquoteright{}, with faster quenches resulting in higher speeds, which may help in two ways.  Firstly, the particles will have less time to accumulate a wetting layer of lutidine, which can affect their contact angle.\cite{White2008}  Secondly, the system spends less time in a state where the particles on the interface can interact, because the interfacial particles reach a jammed state quicker, and this may well allow the particles to better lock in the spinodal structure.\cite{Reeves2015a}

Previous work with the W/L bijel has shown that only certain particle size and quench rate combinations successfully produce bijels\cite{Reeves2015a} \textendash{} the fastest rates are required to form bijels with MPs, but NPs can form bijels at any rate in the range 1 to 350\degree{}C/min.  If we conclude that reducing the quench rate results in a final bijel structure of reduced hyperbolic character, then by applying that reasoning to the 367 nm data point in Figure \ref{quench_rate}(b) we can further illustrate why successful bijel formation with MPs at slower rates is unlikely.  When the area-averaged Gaussian curvature gets closer to the region $\langle K \rangle \Sigma^{-2} \geq 0$, it becomes more and more like a standard Pickering emulsion with increasing amounts of interfacial area curved in the same direction (i.e. droplet morphology), which has been observed previously.\cite{Reeves2015a}  Hence, we can argue that bijel formation will succeed only when the combination of particle size and quench rate results in a (final) structure with sufficiently negative area-averaged Gaussian curvature, and we now know (approximately) what the sufficient value is, and what the range of required parameters is.

\subsection{Curvature change with time}
\begin{figure}[h]
\centering
\begin{subfigure}[b]{0.4\textwidth}
\includegraphics[width=\textwidth]{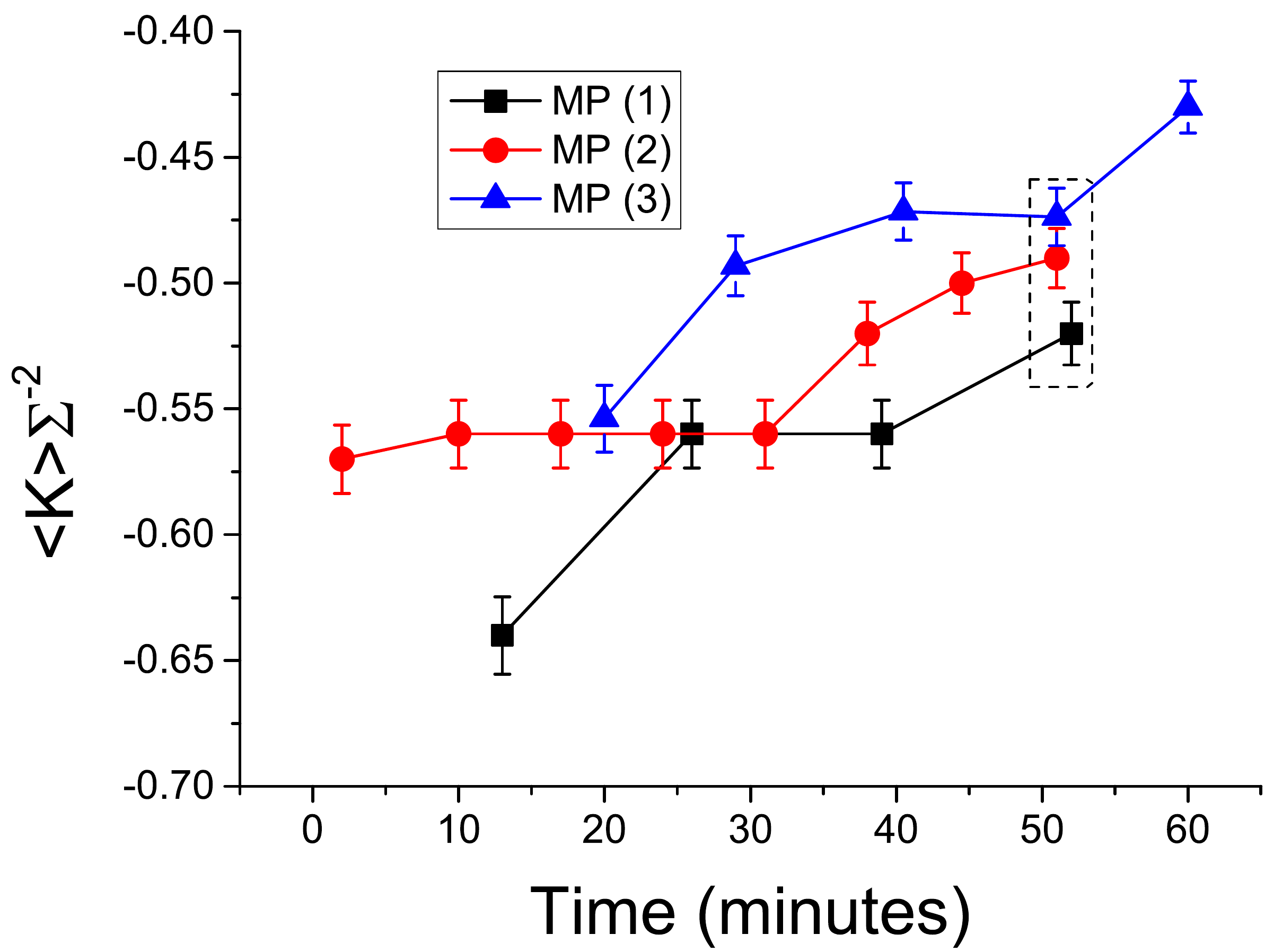}
 \caption{}
\end{subfigure}
\begin{subfigure}[b]{0.4\textwidth}
\includegraphics[width=\textwidth]{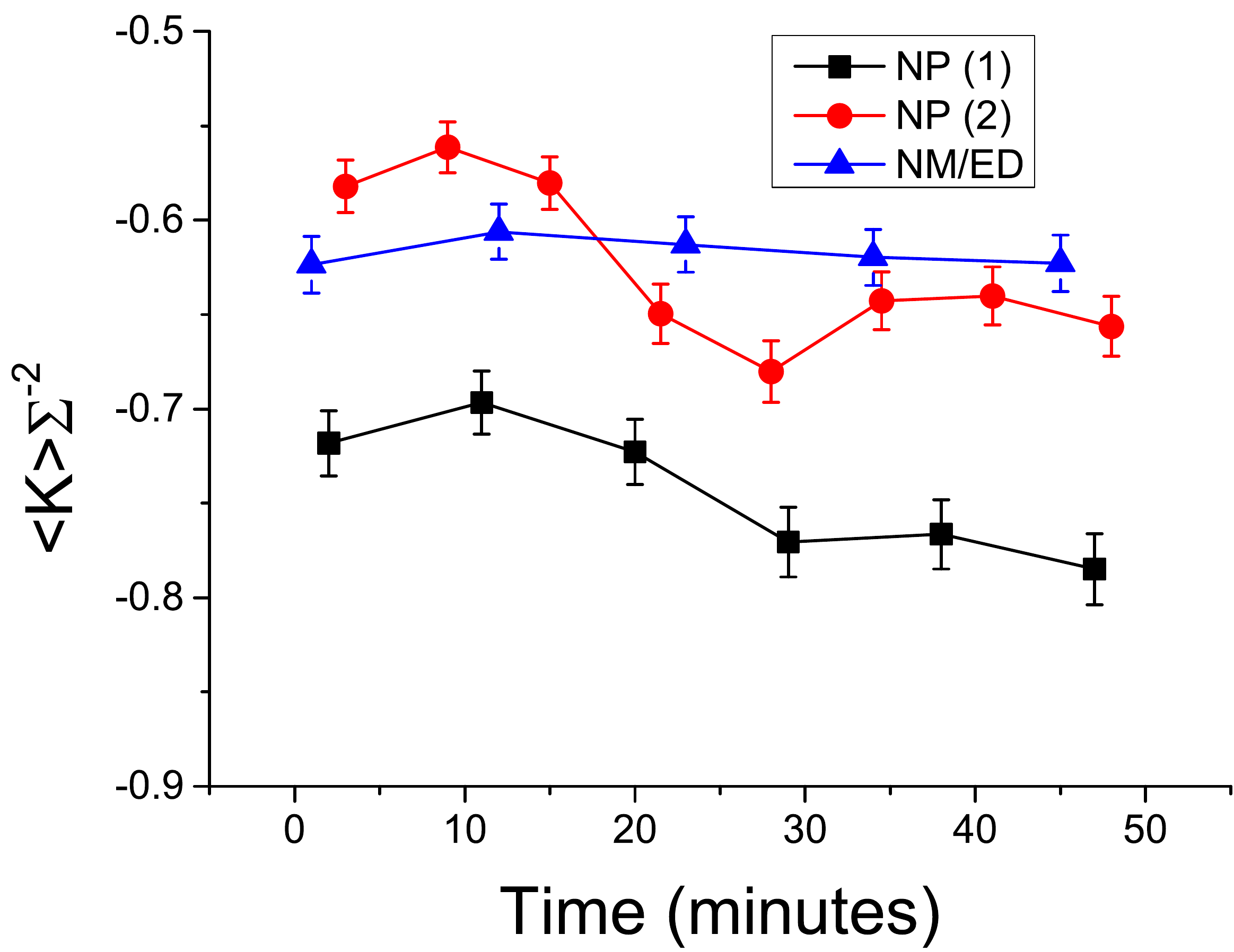}
\caption{}
\end{subfigure}
\caption{(a) The change in area-averaged Gaussian curvature as a function of time for three separately prepared MP bijels (black, red and blue).  The data points in the dashed box are the ones used to generate the MP 350$\degree$C/min data point in Figures \ref{particle_size} and \ref{quench_rate}.  (b) The change in area-averaged Gaussian curvature as a function of time for 2 NP bijels (black and red) and a MP nitromethane/ethanediol bijel (blue, see text for details).  Error bars are $\pm2.4\%$, the error derived from the analysis test in section 2.6.}
\label{Time}
\end{figure}

Figure \ref{Time} shows the area-averaged Gaussian curvatures as a function of time from the onset of phase separation, for three uniquely-prepared (by quenching at 350\degree{}C/min) samples of (a) MP stabilized bijels and (b) (two unique) NP stabilized bijels along with one sample of a bijel made with a different liquid-liquid combination (and MPs), nitromethane/ethanediol (NM/ED).\cite{Tavacoli2011}  The data points in the dashed box in (a) were the ones averaged to create the 367 nm 350\degree{}C/min data point shown in Figures \ref{particle_size}(b) and \ref{quench_rate}(b), as these were the points closest together on the time axis.  In all cases, the values plotted are obtained from the maximum-isosurface area reconstructions, and the error bars are $\pm 2.4 \%$.

In the MP W/L case (Figure \ref{Time}(a)), there is a marked upward movement of the Gaussian curvature in each timeseries, with the final points in all timeseries at least a full error bar above the initial points.  In the NP W/L or the MP NM/ED case however (Figure \ref{Time}(b)), there is no such upward trend, and arguably a downward trend in the case of the NP W/L bijels (red and black).  It is interesting to note that the MP bijels begin their \textquoteleft{}life\textquoteright{} with a structure not too dissimilar from the NP bijels, but mutate over the course of about an hour to end up with a less hyperbolic, sub-optimal structure.

This phenomena is quite unexpected as all of these samples are considered to be in their \textquoteleft{}jammed\textquoteright{} state.  Since the area-averaged Gaussian curvature values have been normalized with respect to the surface-to-volume ratio ($A/V$), such movement cannot be attributed to any slight reduction in ($A/V$), i.e. coarsening (which there is, see the Supplementary Information).  Nor can it be related to a change in genus (and therefore pinch-off events) due to the inability to invoke the Gauss-Bonnet theorem (see Experimental Methods).  Therefore, the movement in Gaussian curvature towards zero indicates a change in the curvature distributions, which we call {\it mutation}.

The movement in Gaussian curvature in the case of MPs can be related with the time over which the elastic modulus of the W/L bijel has been measured to increase,\cite{Lee2012} which can further be associated with the phenomenon of monogelation \textendash{} the emergence of a permanently-bonded particle network which can survive the remixing of the liquids.\cite{Sanz2009}  We observe that a NP-bijel also forms a monogel, but crucially, only requires an incubation period of $<$ 1 min (data not shown here).  The NM/ED bijel does not form a monogel.\cite{Lee2012} Hence, it is possible that the process of gelation of particles at the L-L interface leads to a mutation of the bijel structure.  
\begin{figure}[h]
\centering
\includegraphics[width=0.45\textwidth]{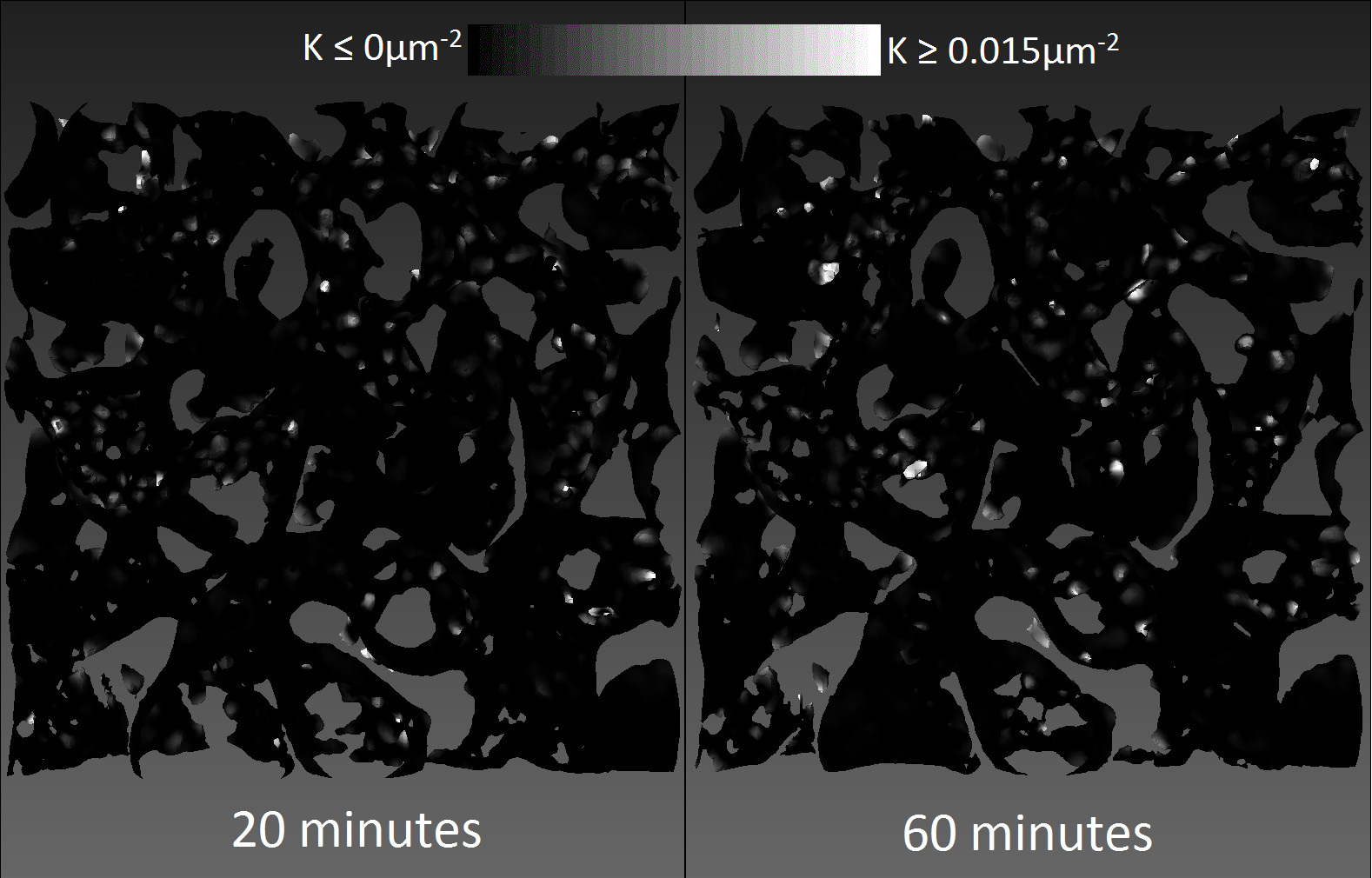}
\caption{A comparison of the isosurfaces representing a 20 minute old bijel and a 60 minute old bijel, stabilized by MPs.  The black areas have Gaussian curvature values less than zero, whereas the bright areas have positive Gaussian curvature, to varying degrees indicated by the brightness.}
\label{comparison}
\end{figure}



In an attempt to ascertain whether the mutation is a global (happens equally everywhere) or local (happens at only specific locations) phenomenon, we have mapped the Gaussian curvature values onto the reconstructed isosurface and compared the 20 min and 60 min-old bijels from the blue timeseries in Figure \ref{Time}(a), shown in Figure \ref{comparison}.  We have thresholded the values so that only positive values appear bright (with varying degrees of brightness), while the black spots are at zero or less.

What we see in Figure 9 are subtle differences in the locations of positive Gaussian curvature, and larger bright spots in the 60 min old sample compared to the 20 min old sample.  The changes are so small that it is difficult to say whether there are specific regions which change more than others.  Nonetheless, it is the entirety of these small changes which results in the shift in the curvature distributions, and hence the metric $\langle K \rangle \Sigma^{-2}$.  The precise mechanism behind the mutation will need to be confirmed by further study.

\section{Conclusions}
In this paper, we have performed an extensive characterization of the topology of bijels produced using different particle sizes and quench rates, and monitored the structures over time.  Optimally hyperbolic (large negative Gaussian curvature) structures are obtained when smaller particles and faster quench rates are used.  Nanospheres appear better able to lock-in the spinodal pattern of the L-L interface during demixing, which corroborates previous work focussing on the specific effect of particle size on the formation of bijels.\cite{Reeves2015a}  Microparticle stabilized bijels appear to mutate away from an optimal structure, which may be related to the monogelation phenomenon.\cite{Sanz2009}  Nanoparticle stabilized bijels, which form a monogel on much shorter timescales, become marginally more optimal over time, and a non-monogelling system (nitromethane/ethanediol) shows no such mutation.  In practical terms, we now have a characterization which enables us to systematically optimize the morphology of (final-state) bijels, which will assist in its development for applications.

\section{Acknowledgements}
M.R. is grateful to EPSRC for funding his PhD studentship.  J.H.J.T acknowledges the Royal Society of Edinburgh/BP Trust Fellowship, and the University of Edinburgh for support through the Chancellor's Fellowship.  This work has been financially supported by EPSRC grant number EP/J007404.  The authors would like to thank Rob Wallace for assistance with X-ray CT acquisition, Andrew Schofield for particle synthesis, Katherine Rumble for the preparation of NM/ED bijels and Mike Cates, Aidan Brown, Joe Tavacoli and Michiel Hermes for useful discussions.





\bibliography{curvature} 
\bibliographystyle{rsc} 

\end{document}


\maketitle
\section{Errors}
We have noticed that the threshold value used to create the isosurface most closely resembling the raw data (assessed by visual inspection) was the one which also maximized the area of the isosurface.  For comparing curvatures across a range of samples therefore, the threshold value for the maximum area was found, and those curvature values recorded.  To take account of sampling error, the entire procedure (from sample preparation, to data acquisition, to image analysis) was repeated three times, and an average of quantities taken.  Where repeat sampling was not feasible (e.g.  not enough identical material being available) the threshold value was varied around that which gave the maximum area, and an average of those curvature values taken.  Hence, the error bars on these data points represent the thresholding error in the measurement of one sample.  It is made clear in the text which data points have been multiply sampled and which have not.  Also, the thresholding error was found to decrease when using a 40$\times$ objective rather than a 20$\times$ objective, but that the absolute curvature values were not affected \textendash{} see Supplementary Table 1.

Another source of error is due to the \textquoteleft{}fishtank\textquoteright{} effect, whereby the position of the focal plane moves either more or less than the nominal Z increment depending on the relative refractive indices of the sample medium and the medium surrounding the collection optics.  In this case, by using the average refractive index of the water/lutidine mixture\cite{Grattoni1993}($n_{WL} \approx 1.4$), and the refractive index of the collection medium (air, $n_A \approx 1$) the correction is calculated to be $n_{WL}/n_{A} = $ 1.4 times the nominal Z difference.\cite{Fishtank}  Therefore, the analysis has also been tested on the stretched data, by modifying the voxel sizes specified when loading the data into Avizo.  We found that the fishtank effect, while slightly modifying the absolute numbers, did not significantly affect the trends in the data \textendash{} see Supplementary Figures 2 and 3.
\begin{supptable}[h]
\centering
\begin{tabular}{|c|c|c|c|}
\hline
\textbf{Location} & \textbf{Magnification} & \textbf{$\langle K \rangle \Sigma^{-2}$} & $\pm$ \\
\hline
A & 20$\times$ & -0.50 & 0.06 \\
\hline
B & 20$\times$ & -0.49 & 0.08 \\
\hline
C & 40$\times$ & -0.50 & 0.01 \\
\hline
D & 40$\times$ & -0.49 & 0.01 \\
\hline

\end{tabular}
\caption{The results of a curvature analysis performed on a nanoparticle stabilized bijel quenched at 350\degree{}C/min, sampled at four different positions (A-D) and using two different objectives (20$\times$ and 40$\times$).  Samples showed good internal consistency, with the only effect of changing objective being a reduced thresholding error.}
\end{supptable}

\begin{suppfigure}[h]
\centering
\includegraphics[width=\textwidth]{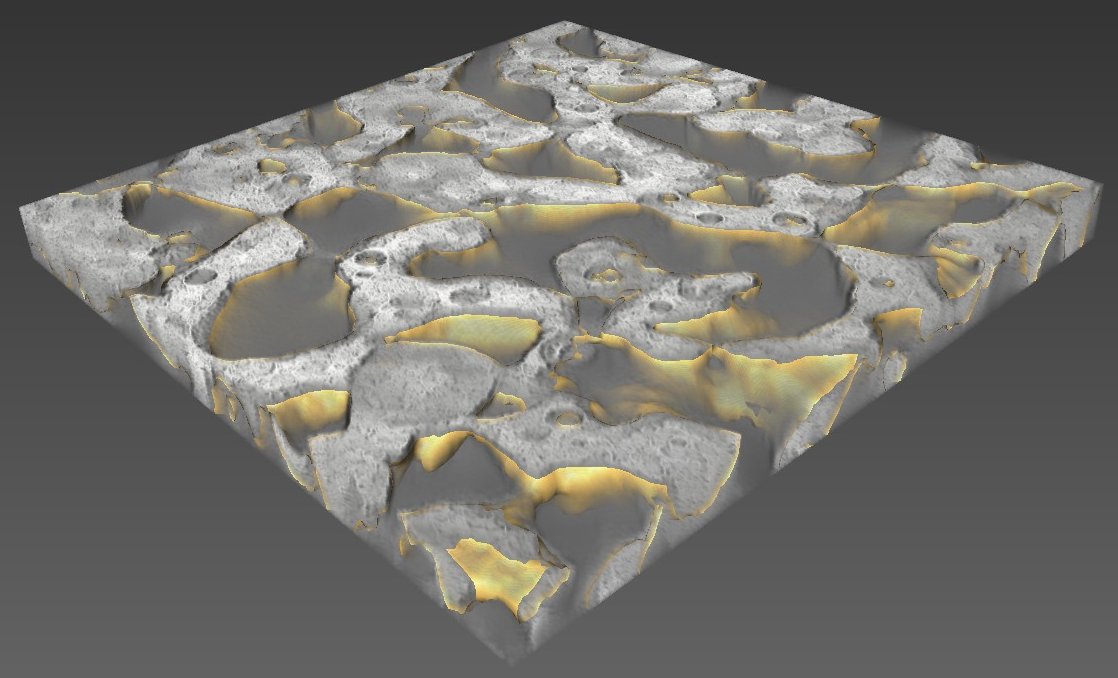}
\caption{A 3D volume rendering of a microparticle stabilized bijel (greyscale) with the calculated isosurface superimposed (yellow).  The isosurface clearly follows the pattern of the raw data, meaning that the curvature analysis performed on the isosurface can be said to represent the curvature values of the raw data, i.e. the bijel.}
\end{suppfigure}

\begin{suppfigure}[h]
\centering
\begin{subfigure}[b]{\textwidth}
\includegraphics[width=\textwidth]{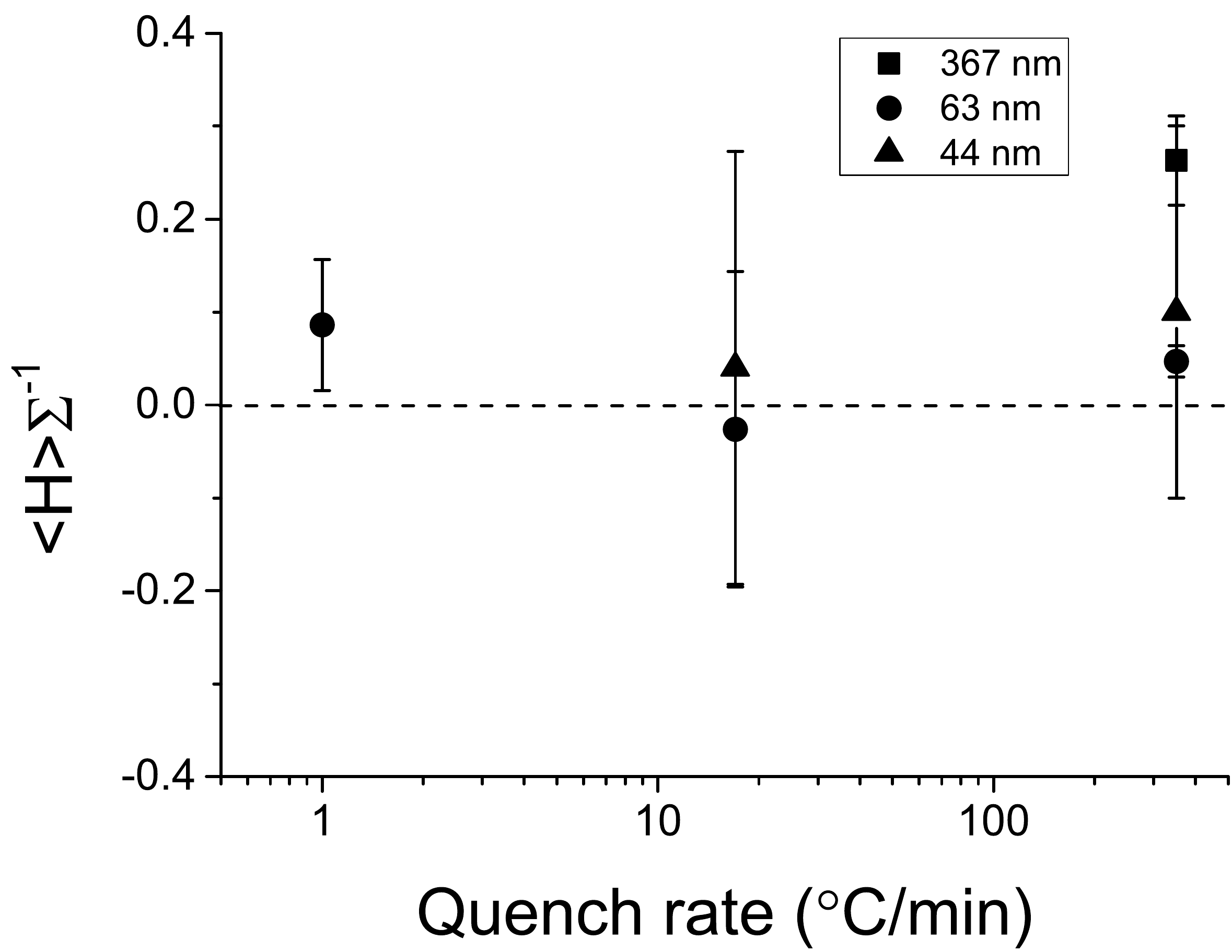}
\caption{}
\end{subfigure}
\begin{subfigure}[b]{\textwidth}
\includegraphics[width=\textwidth]{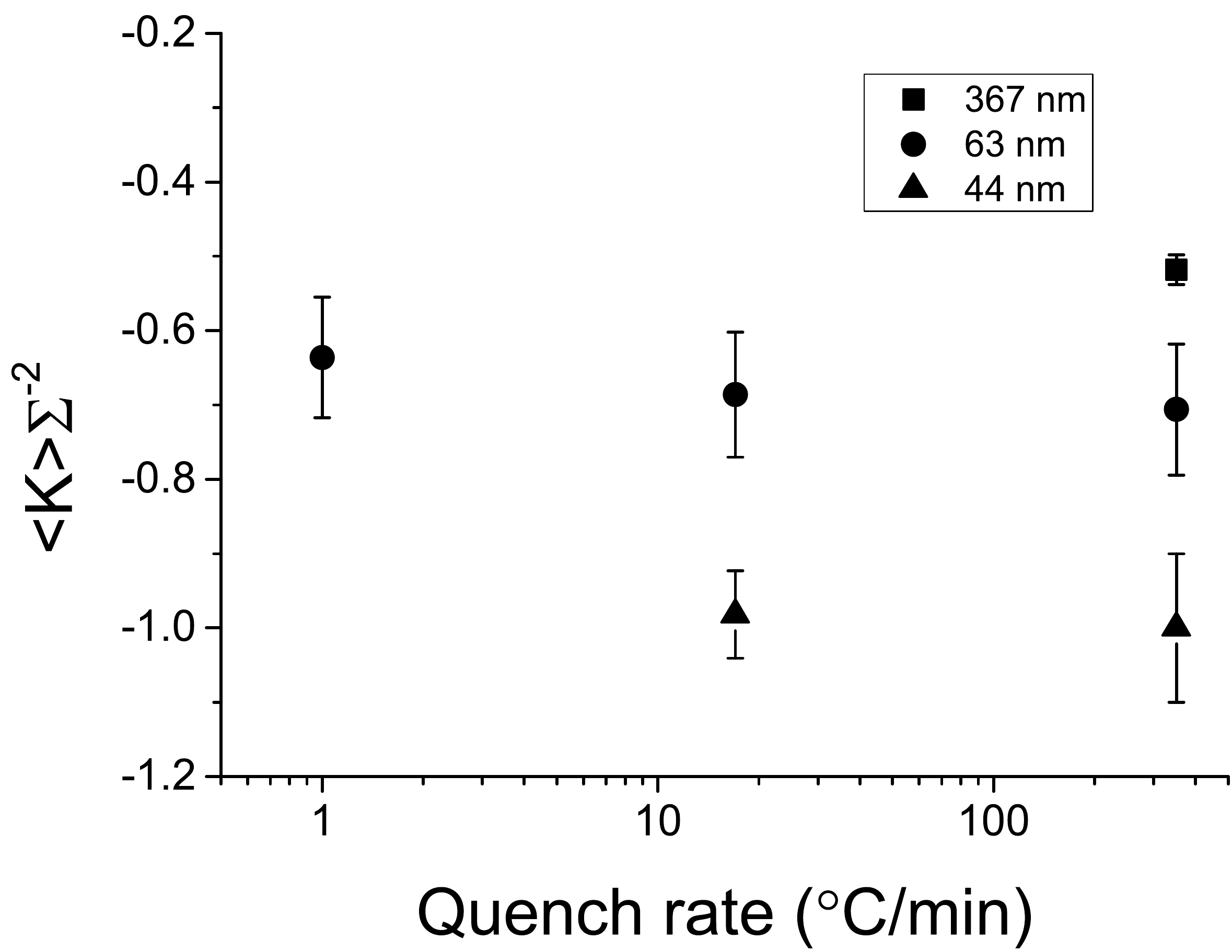}
\caption{}
\end{subfigure}
\caption{The area-averaged mean curvatures (a) and Gaussian curvatures (b) as a function of quench rate, with the three sizes of particles used in the study, after the data has been corrected for the fishtank effect.  The absolute values are slightly changed, but the trends remain.  This analysis was not used in the main paper because of the non-linear effect of stretching the voxels on the accuracy of the curvature measurement protocol, as evidenced in Supplementary Figure \ref{gyroid_lambda}.}
\end{suppfigure}

\begin{suppfigure}[h]
\centering
\includegraphics[width=\textwidth]{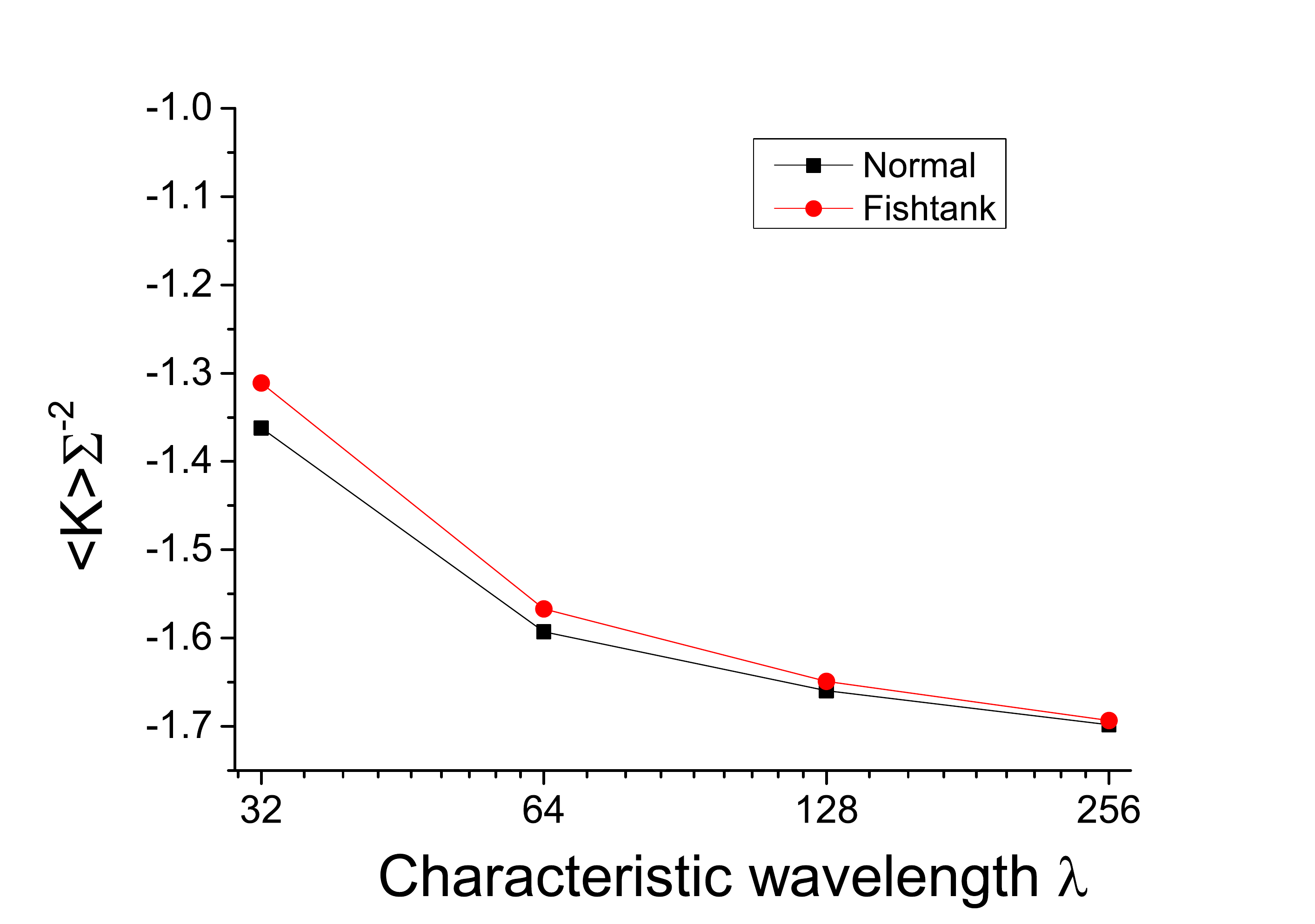}
\caption{The area-averaged Gaussian curvature of a simulated ideal gyroid, in a volume of 512 by 512 by 128 pixels, as a function of the gyroid wavelength (black), and the same after correcting for the fishtank effect (red).  As the wavelength is reduced, the curvature measurement protocol becomes less accurate, i.e. produces a result further from the theoretical result expected ($\approx -1.7$).  Also, the fishtank effect becomes more prominent as the wavelength is reduced.  However, our data lies in the range between $\lambda$=128 and $\lambda$=256 pixels, where the curvature analysis protocol is hardly affected by taking into account the fishtank effect.}
\label{gyroid_lambda}
\end{suppfigure}

\begin{suppfigure}[h]
\centering
\includegraphics[width=\textwidth]{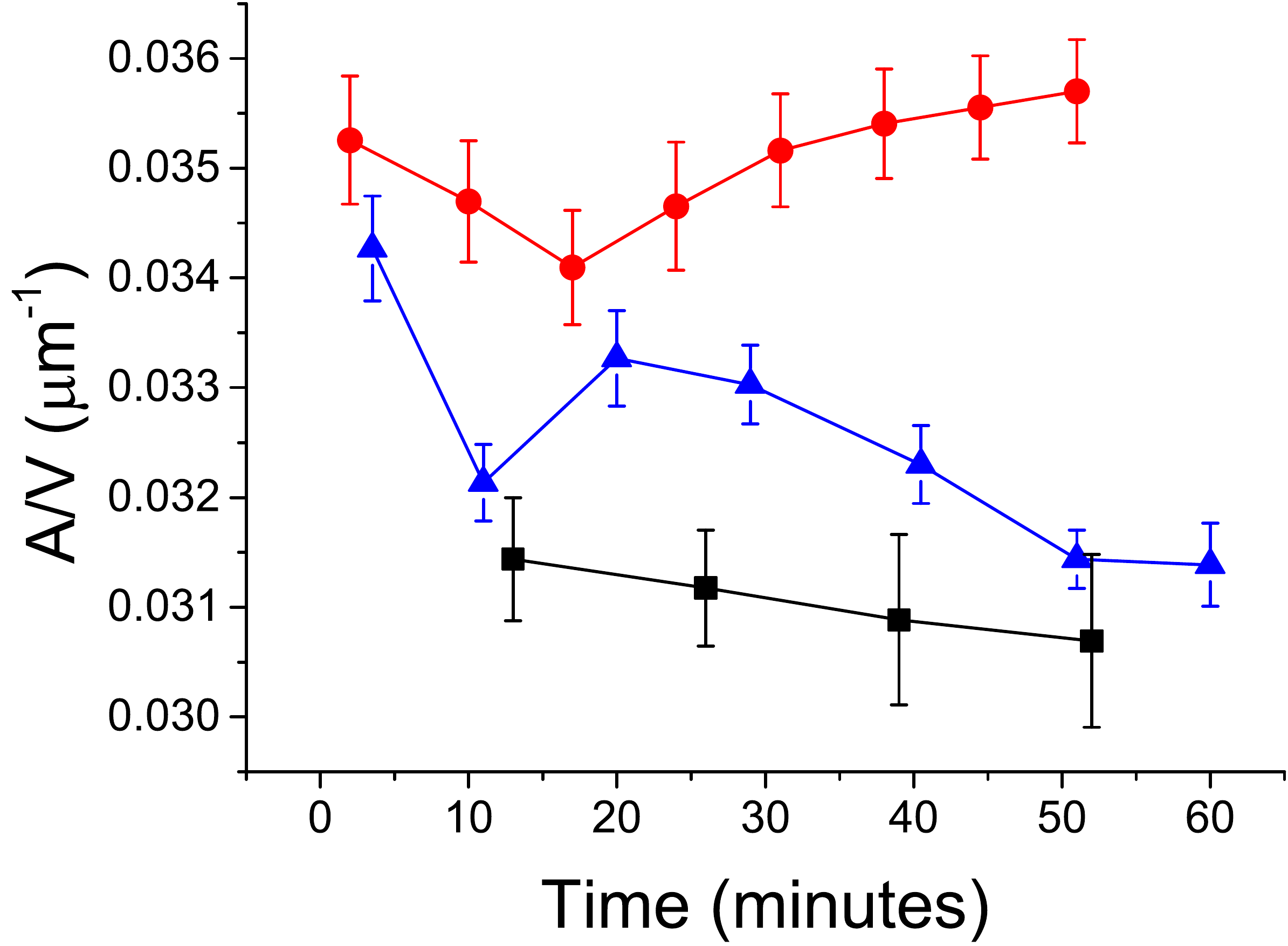}
\caption{The change in the surface-to-volume ratio ($A/V$) as a function of time from the onset of phase separation, in the case of three separately prepared microparticle-stabilized bijels: the same ones as presented in Figure 8 (a) of the main text.  A reduction in $A/V$ indicates coarsening of the bicontinuous structure \textendash{} this is observed in 2 out of 3 of the data sets.  The behaviour of the red data set can be accounted for by a slight drift in the bijel sample relative to the data volume, meaning that this data set is a less reliable measure of any time dependence in $A/V$.  Nonetheless, any change in area-averaged curvature cannot be explained by coarsening, since the curvatures are normalized with respect to $A/V$.}
\end{suppfigure}
\clearpage
\bibliography{curvature}
\bibliographystyle{rsc}